\documentclass[aps,prd,showpacs,amssymb,floatfix,nofootinbib,superscriptaddress]{revtex4-1}
\usepackage{graphicx,amsmath,subfigure}

\newcommand{\eeq}{\end{equation}}
\newcommand{\bea}{\begin{eqnarray}}

\def\ltsima{$\; \buildrel < \over \sim \;$}
\def\simlt{\lower.5ex\hbox{\ltsima}}
\def\gtsima{$\; \buildrel > \over \sim \;$}
\def\simgt{\lower.5ex\hbox{\gtsima}}

\newcommand{\hGR}{\mathbf{h}_\mathrm{GR}}
\newcommand{\hAP}{\mathbf{h}_\mathrm{AP}}
\newcommand{\tbf}{\theta}
\newcommand{\ttr}{\hat \theta}

\def\lesssim{\mathrel{\hbox{\rlap{\hbox{\lower4pt\hbox{$\sim$}}}\hbox{$<$}}}}
\def\gtrsim{\mathrel{\hbox{\rlap{\hbox{\lower4pt\hbox{$\sim$}}}\hbox{$>$}}}}
\def\alt{\mathrel{\hbox{\rlap{\hbox{\lower4pt\hbox{$\sim$}}}\hbox{$<$}}}}
\def\agt{\mathrel{\hbox{\rlap{\hbox{\lower4pt\hbox{$\sim$}}}\hbox{$>$}}}}
\def\PRD{{\it Phys. Rev.} D~}
\def\PRL{{\it Phys.Rev.} Lett~}

\def\CQG{{\it Class. Quantum Grav.}}

\def\gta{\ifmmode {\mathbin{\lower 3pt\hbox   
    {$\,\rlap{\raise 5pt\hbox{$\char'076$}}\mathchar"7218\,$}}}
    \else {${\mathbin{\lower 3pt\hbox
    {$\rlap{\raise 5pt\hbox{$\char'076$}}\mathchar"7218\,$}}}
    $}\fi}
\def\lta{\ifmmode {\,\mathbin{\lower 3pt\hbox   
    {$\,\rlap{\raise 5pt\hbox{$\char'074$}}\mathchar"7218\,$}}}
    \else {${\mathbin{\lower 3pt\hbox
    {$\rlap{\raise 5pt\hbox{$\char'074$}}\mathchar"7218\,$}}}
    $}\fi}

\newcommand{\SU}{\affiliation{Department of Physics, Syracuse University, Syracuse, NY 13244, USA.}}
\newcommand{\CU}{\affiliation{Institute of Astronomy, Madingley Road, CB3 0HA Cambridge, UK.}}

\begin{document}
\title{Importance of including small body spin effects in the modelling of intermediate mass-ratio inspirals. II Accurate parameter extraction of strong sources using higher-order spin effects}

\author{E.A. Huerta}\email{eahuerta@syr.edu}\SU\CU
\author{Jonathan R. Gair}\email{jgair@ast.cam.ac.uk}\CU
\author{Duncan A. Brown}\email{dabrown@phy.syr.edu}\SU


\date{\today}

\begin{abstract}        
We improve the numerical kludge waveform model introduced in Huerta \& Gair (2011)~\cite{smallbody} in two ways. We  extend the equations of motion for spinning black hole binaries derived by Saijo et al. \cite{maeda} using spin--orbit and spin--spin couplings taken from perturbative and post--Newtonian (PN) calculations at the highest order available. We also include first-order conservative self--force corrections for spin-orbit and spin-spin couplings, which are derived by comparison to PN results. We generate the inspiral evolution using fluxes that include the most recent calculations of small body spin corrections, spin-spin and spin-orbit couplings and higher-order fits to solutions of the Teukolsky equation. Using a simplified version of this model in \cite{smallbody}, we found that small body spin effects could be measured through gravitational wave observations from intermediate-mass ratio inspirals (IMRIs) with mass ratio \(\eta \gtrsim10^{-3}\), when both binary components are rapidly rotating. In this paper we present results of Monte Carlo simulations of parameter estimation errors to study in detail how the spin of the small/big body affects parameter measurement using a variety of mass and spin combinations for typical  IMRI  sources. We have found that for IMRI events involving a moderately rotating intermediate mass black hole (IMBH) of mass  \(10^4M_{\odot}\) and a rapidly rotating central supermassive black hole (SMBH) of mass  \(10^6M_{\odot}\), gravitational wave observations made with LISA at a signal-to-noise ratio (SNR) of \(1000\) should be able to determine the inspiralling IMBH mass, the central SMBH mass, the SMBH spin magnitude,  and the IMBH spin magnitude to within fractional errors of \(\sim 10^{-3},\,10^{-3},\,10^{-4}\),  and \(10^{-1}\), respectively. LISA should also be able to determine the location of the source in the sky and the SMBH spin orientation to within \(\sim 10^{-4}\) steradians. Furthermore, we show that by including conservative corrections up to 2.5PN order, systematic errors no longer dominate over statistical errors. This shows that search templates that include small body spin effects in the equations of motion up to 2.5PN order should allow us to perform accurate parameter extraction for IMRIs with typical SNR\(\,\sim1000\). 
\end{abstract}


\maketitle

\section{Introduction}    

Developing waveform templates that accurately model the inspiral of stellar mass compact objects (CO) into supermassive black holes (SMBHs) has enabled us to shed light on the scientific payoffs that would be obtained if we were able to detect and characterize  a large number of these events. Current estimates suggest that if a space-based detector like LISA~\cite{noise}  made a single detection of these extreme-mass-ratio inspirals (EMRIs), we could determine the intrinsic parameters of the system, i.e., the masses of the binary and the spin of the central SMBH, to a fractional accuracy of \(\sim10^{-4}\), the orientation of the spin of the SMBH and the position of the source in the sky to a resolution of \(\sim 10^{-3}\) steradians and the luminosity distance to an accuracy of \(\sim10\%\) \cite{cutler,cons}. 

Most existing EMRI models ignore the spin of the smaller object and so recent work on source modelling of EMRIs has tried to determine the impact of ignoring the small body spin effects on source detection and parameter estimation. Barack \& Cutler estimated  that over the last year of inspiral, the inclusion of small body spin effects would affect the accumulated orbital phase by a few radians~\cite{cutler}.  The reason being the dimensionless spin parameter of the inspiralling CO is multiplied by a factor of \(\eta\) in the equations of motion \cite{maeda}, and hence for EMRIs of mass ratio \(\eta\sim10^{-5}\), the contribution of the CO spin to the binary's dynamics is highly suppressed. Furthermore, Burko \cite{burko} has shown that omitting local self-force corrections in the modeling of spinning COs that inspiral into supermassive Schwarzschild BHs is not needed to accurately model the orbital evolution, and that such omission will only decrease our ability to determine the spin rate of the companion. 

We have recently explored the effect of spin using a different approach. We developed a waveform model that includes small body spin corrections and determined  the mass--ratio threshold \(\eta\) at which the small body spin effects become relevant for the dynamics. We found that the spin of the inspiralling CO could be accurately determined from gravitational wave (GW) observations only for \(\eta\gtrsim10^{-3}\). We also showed that, in this regime, small body spin corrections become particularly important when both components of the binary are rapidly rotating.

A mass ratio of $\eta \sim 10^{-3}$ can no longer be described as an EMRI, but represents an intermediate-mass-ratio inspiral (IMRI). This distinction is a modelling one. For small mass ratios, $\eta \lesssim 10^{-4}$ (EMRIs), the gravitational waveforms can be modelled accurately using black hole perturbation theory (BHPT), while for near equal-mass systems, $\eta \gtrsim 10^{-1}$, the waveforms can be computed using post-Newtonian (PN) techniques. IMRIs lie between these two limits, in a regime in which the velocities of the binary components are too large for PN theory to apply, but the mass ratio is sufficiently large that perturbation theory cannot be used.

Our findings in~\cite{smallbody} suggested that search templates that aim to accurately measure the properties of IMRIs would also have to include small body spin effects. There are two different types of system that could be observed as IMRIs and both of these involve intermediate mass black holes (IMBHs) with mass $M\sim10^2$--$10^4M_{\odot}$. An IMBH that inspirals into a SMBH in the centre of a galaxy could be detected by LISA if the SMBH has mass in the appropriate range $\sim10^4$--$10^6M_{\odot}$. Similarly, if a stellar mass compact object, a neutron star or black hole, inspirals into an IMBH of mass $\sim10^2-10^3M_{\odot}$ the emitted GWs might be observed by ground-based detectors such as Advanced LIGO or the Einstein Telescope~\cite{man,etgair,firstpaper,thirdpaper}. Such events could occur in the dense stellar environments of globular clusters that contain IMBHs. Both types of system constitute IMRIs, but in this paper we focus on the former category.

For LISA IMRIs, an accurate determination of the spin of the inspiralling CO will provide information about the formation channels of IMBHs.  Numerical studies suggest that high observed spins would imply that much of the mass of these objects was obtained by gas accretion from a disc, whereas moderate spins would favour a scenario in which the IMBH formed as a result of a major merger between comparable-mass black holes, and low spins would imply that the IMBH acquired its mass through a series of minor mergers with smaller objects \cite{seoane,mandel}.  Therefore, it is important to determine if templates that include small body spin effects at the highest order available will allow us to measure the spin distribution of IMBHs through IMRI observations. In this paper we extend the results of~\cite{smallbody} by including higher order corrections in the waveform model and exploring a wider variety of possible systems.

To perform this analysis, we will make use of an improved version of the ``numerical kludge'' waveform scheme. In this framework, one combines an accurate prescription for the generation of the inspiral trajectory with a pseudo-flat spacetime wave generation formula. Even though this construction is inconsistent, it has been shown that numerical kludge waveforms capture the main features of Teukolsky-based waveforms \cite{kludge}.  In our improved scheme, we extend the spinning black hole binary model derived by Saijo et al~\cite{maeda} by: (i) generating the inspiral trajectory using fluxes of energy and angular momentum that include BHPT calculations for spin-spin and spin-orbit couplings at the highest order available, and (ii) by amending the orbital phase evolution with conservative self--force corrections for spin--orbit and spin--spin couplings using the same method that we used to  compute perturbative conservative corrections in~\cite{cons,smallbody}.   

The analysis carried out in \cite{volon} suggests that the spin distribution of massive Kerr BHs is heavily skewed toward rapid spins below redshift \(z\sim 5\). This analysis suggests that gravitational radiation generated by IMRI sources involving supermassive Kerr BHs may include small body spin effects. Hence, it is worthwhile carrying a detailed study of the influence of higher-order spin corrections for parameter estimation of IMRIs.  For completeness in our analysis, we also estimate the systematic errors that arise when we omit higher-order corrections in the waveform template. We find that by including the conservative corrections up to 2.5PN order in search templates, it should be possible to reduce the magnitude of systematic errors so that  they do not dominate over the parameter errors arising from instrumental noise, even for strong sources. 

This paper is organized as follows. In Section~\ref{s2} we describe our updated kludge waveform template that includes the most recent results from both perturbative and PN calculations for spin-spin and spin-orbit calculations in a physically consistent manner. In section~\ref{s3} we summarize the wave generation model used in our studies. In Section~\ref{s4} we describe the Fisher Matrix formalism used to estimate parameter estimation errors. We use this formalism in Section~\ref{s5} to explore how the accuracy of parameter measurement depends on the spin of the small/big body using a representative sample of binary systems and a variety of spin configurations. In Section~\ref{s6} we estimate the systematic errors  that arise from omitting higher order corrections in the waveform template using the formalism introduced in~\cite{vallisneri}.  We summarize our findings in Section~\ref{s7}.

\section{Implementation of higher-order spin effects in the equations of motion of spinning black hole binaries}
\label{s2}

In the first paper of this series \cite{smallbody}, we showed how to extend the kludge waveform model introduced in \cite{kludge} by including small body spin corrections. In this Section, we briefly review this scheme and show how we can improve this model by including perturbative and PN corrections for spin-spin and spin-orbit couplings at the highest order available. Thereafter, we will augment the search template with conservative self-force corrections which we derive by comparison to PN results. The aim of this analysis is to provide a search template that includes the most recent calculations, both from perturbative and PN analysis, to shed light on the capabilities of future low-frequency detectors to measure the spin of IMBHs that inspiral into SMBHs, and to find out whether including conservative self-force corrections for spin-spin and spin-orbit couplings at the highest order available will be enough to prevent model errors to dominate over statistical errors.

The equations of motion we shall use to build the IMRI numerical kludge waveform model are those derived in~\cite{maeda}, which describe the orbital evolution of a spinning CO that inspirals into a spinning SMBH.   We shall assume that the spin angular momentum  \({\bf S}_1\) of the inspiralling CO is aligned with the central supermassive Kerr BH spin \({\bf S}_2\) and the orbital angular momentum $L_z$. In this configuration, the orbital evolution can be modeled using Eqs. (1)-(9) of \cite{smallbody}.

The first improvement to the IMRI waveform model introduced in \cite{smallbody} affects the generation of the inspiral trajectory of the spinning CO.  In this analysis, we will use the radiation fluxes  derived by Gair \& Glampedakis~\cite{improved}, augmented  with accurate BHPT results that include small body spin corrections at the highest order available~\cite{ tanaka}. The expressions for energy and angular momentum flux become

\begin{eqnarray}
\dot{E} &=& -\frac{32}{5} \frac{\mu^2}{M^2 }\left(\frac{1}{r}\right)^{5}\Bigg\{ 1 - \frac{1247}{336}\left(\frac{1}{r}\right)+ \left(4\pi-  \frac{73}{12}q-\frac{25}{4}\eta\chi \right)\left(\frac{1}{r}\right)^{3/2} + \left( - \frac{44711}{9072} + \frac{33}{16}q^2 + \frac{71}{8} q\eta\chi\right)\left(\frac{1}{r}\right)^2 \nonumber \\ && +\left(-\frac{8191}{672}\pi + \frac{3749}{336}q + \frac{2403}{112}\eta\chi\right)\left(\frac{1}{r}\right)^{5/2} \, +\textrm{higher order Teukolsky fits}\Bigg\}, \nonumber\\
\dot{J}_z&=& -\frac{32}{5} \frac{\mu^2}{M} \left(\frac{1}{r}\right)^{7/2} \Bigg\{ 1 - \frac{1247}{336}\left(\frac{1}{r}\right)+ \left(4\pi-  \frac{61}{12}q-\frac{19}{4}\eta\chi \right)\left(\frac{1}{r}\right)^{3/2} + \left( - \frac{44711}{9072} + \frac{33}{16}q^2 + \frac{59}{8} q\eta \chi\right)\left(\frac{1}{r}\right)^2 \nonumber \\ && + \left(-\frac{8191}{672}\pi + \frac{417}{56}q + \frac{3559}{224}\eta\chi\right)\left(\frac{1}{r}\right)^{5/2} \, +\textrm{higher order Teukolsky fits}\Bigg\},
\label{new_Ldot}
\end{eqnarray}

 \noindent where \(r=p/M\),  and \(p\) stands for the Boyer-Lindquist radial coordinate, \(\hat s= s/M =\eta \chi\), where \(\chi\) stands for the dimensionless spin parameter of the inspiralling CO, \(\eta=\mu/M\) represents the mass-ratio, and \(q\)  is the dimensionless spin parameter of the central SMBH. 
 
 It is worth pointing out that to derive the conservative self-force corrections we will make use of the terms explicitly included in the expressions for \(\dot{E}\) and \(\dot{J}_z\) in Eq.~\eqref{new_Ldot}. However, the waveforms will be generated including the higher-order Teukolsky fits derived in \cite{improved}. 
 
 Furthermore, as shown by Tanaka et al.~\cite{tanaka}, to evolve circular equatorial orbits for a spinning particle in a Kerr background we can use the  ``circular goes to circular'' rule of~\cite{ori}, i.e., \cite{tanaka}
 
 \begin{equation}
\dot E_{\rm }(r) = \pm \frac{1}{r^{3/2} \pm q }\left(1-\frac{3}{2}\eta\chi\frac{\pm \sqrt{r}-q}{r^2 \pm q \sqrt{r}}\right)  \dot J_{z}(r )=\Omega\dot J_{z}(r ),
\label{cirsp}
\end{equation}

\noindent where \( \mathrm{d} \phi / \mathrm{d} t = \Omega(r)\) stands for the orbital frequency. We should bear in mind that this assumption is consistent with the energy and angular momentum loss rates at linear order in the spin of the particle \cite{tanaka}. 

 These considerations then allow us to evolve the radial coordinate as follows
 
 \begin{equation}
\label{6}
\dot r= \frac{\mathrm{d} r}{\mathrm{d} E}\dot E= \frac{\mathrm{d} r}{\mathrm{d} J_z}\dot J_z .
\end{equation}

\noindent We will generate the inspiral trajectory, \(r(t)\), using the expression for the angular momentum quoted in \eqref{new_Ldot} which includes higher-order spin corrections and higher-order fits toTeukolsky data.  

The search template we have developed so far includes radiative self-force corrections at the highest order available. Hence, for completeness in the modeling, we shall now include conservative self-force corrections. Because the conservative piece of the self-force accumulates and affects the phasing of the waveform over time, we need to include this effect as it could lead to several cycles of phase discrepancy in the kludge model over the inspiral.  We take into account this effect by rewriting the orbital frequency \(\Omega\) in the following form,

\begin{equation}
\frac{\mathrm{d}\phi}{\mathrm{d}t} = \left(\frac{\mathrm{d}\phi}{\mathrm{d}t}\right)_{\mathrm{geodesic}}\bigg(1+  \delta \Omega \bigg),
\label{7.1}
\end{equation}

\noindent where the first term on the righthand side of Eq.~\eqref{7.1} is implicitly defined in Eq.~\eqref{cirsp}, and \(\delta \Omega\) is a frequency shift which depends on the instantaneous orbital parameters, and which is still unknown for spinning particles that inspiral into Kerr SMBHs. At present, the self-force program has been successful in computing gravitational self-force data for bound eccentric geodesics in Schwarzschild space-time, and in developing numerical algorithms to calculate the self-force on a scalar charge moving in bound eccentric equatorial geodesics in Kerr space-time  \cite{sagoec,sago,warleor,war,warac}. The extension of these numerical algorithms from the scalar to the gravitational case is the main goal of the self-force community  in the foreseeable future. Hence, for the time being,  the computation of gravitational self-force data for spinning particles in Kerr space-time is beyond the scope of the self-force program. However, we can still make progress by using the conservative self-force corrections that have been derived within the PN formalism. In order to be consistent with the expressions for the radiative part of the self-force, see Eq.~\eqref{new_Ldot}, we will use conservative PN self-force corrections up to 2.5PN order. 

To implement PN conservative self-force calculations in the kludge waveform, we will modify the kludge orbital frequency and its first time derivative. We will then compare these asymptotic observables with their PN counterparts. This comparison will enable us to determine coordinates between both formalisms, and compute the conservative pieces. We shall start by expanding Eq.~\eqref{cirsp} as follows 

\begin{equation}
\label{eq.5}
\Omega = \frac{1}{M}\left(\frac{1}{r}\right)^{3/2}\left(1- \left(q + \frac{3}{2}\eta\chi\right)\left(\frac{1}{r}\right)^{3/2} + \frac{3}{2} q\eta\chi\left(\frac{1}{r}\right)^{2} + O\left(\frac{1}{r}\right)^{3}\right),
\end{equation}

\noindent which takes the following form when we include conservative corrections 

\begin{eqnarray}
\frac{\mathrm{d}\phi}{\mathrm{d}t} \equiv \Omega &=& \frac{1}{M}\left(\frac{1}{r}\right)^{3/2}\left(1- \left(q + \frac{3}{2}\eta\chi\right)\left(\frac{1}{r}\right)^{3/2} + \frac{3}{2} q\eta\chi\left(\frac{1}{r}\right)^{2} \right)\bigg(1 + \delta \Omega \bigg),  \nonumber\\ 
 & =& \frac{1}{M}\left(\frac{1}{r}\right)^{3/2}\left(1- \left(q + \frac{3}{2}\eta\chi\right)\left(\frac{1}{r}\right)^{3/2} + \frac{3}{2} q\eta\chi\left(\frac{1}{r}\right)^{2} \right)\Bigg\{1 +  \nonumber\\ &+&\eta \left( c_0 + c_2 \left(\frac{1}{r}\right)+ (c_{3} + q\, c_{3.1} +\chi\, c_{3.2} )\left(\frac{1}{r}\right)^{3/2} + (c_4 + c_{4.1}\,q\chi)\left(\frac{1}{r}\right)^{2} + (c_{5} + q\, c_{5.1} +\chi\, c_{5.2} )\left(\frac{1}{r}\right)^{5/2}\right)\Bigg\}.\nonumber\\
 \label{omCCone}
\end{eqnarray}

To modify the first order time derivative of the orbital frequency \eqref{omCCone}, we will use 

\begin{eqnarray}
\frac{{\mathrm d}\Omega}{{\mathrm d}t}&=& \frac{{\mathrm d}\Omega}{{\mathrm d} r} \frac{\mathrm{d}r}{\mathrm{d}t}, \quad {\rm with}
\label{omdot}\\
\frac{\mathrm{d}r}{\mathrm{d}t} &=& - \frac{64}{5}\frac{\eta}{M} \left(\frac{1}{r}\right)^{3}\Bigg\{1- \frac{743}{336}\left(\frac{1}{r}\right) + \left(4 \pi - \frac{133}{12} q -\frac{35}{4}\, \eta\chi\right)\left(\frac{1}{r}\right)^{3/2} \nonumber\\ &+& \left(\frac{34103}{18144} + \frac{81}{16} q^{2} + \frac{95 \, q\eta\chi}{8}\right)\left(\frac{1}{r}\right)^{2} +  \left(-\frac{4159}{672} \pi - \frac{1451}{56} q -\frac{1271}{672}\, \eta\chi\right)\left(\frac{1}{r}\right)^{5/2}\Bigg\}.
\label{rdot}
\end{eqnarray}

\noindent Note that for consistency with Eq.~\eqref{omCCone}, we have used \(\dot{r}\) at 2.5PN order. 

To compute the first time derivative of the orbital frequency within the kludge formalism, \(\dot{\Omega}\), we assume that the time evolution of the radial coordinate is given by Eq.~\eqref{rdot}, and choose a gauge in which the \(\eta^2\) terms in \(\dot{r}(t)\) that are not proportional to \(\chi\) vanish. Under these assumptions, the first time derivative of  \eqref{omCCone} is given by

\begin{eqnarray}
\frac{\mathrm{d} \Omega}{ \mathrm{d} t} &=& \frac{96}{5}\frac{\eta}{M^2}\left(\frac{1}{r}\right)^{11/2}\Bigg\{ 1+ \eta\, c_0 + \frac{1}{r} \left(-\frac{743}{336} + \eta \left[\frac{5}{3} c_2 - \frac{743}{336} c_0 \right]\right) + \nonumber\\  &+& \left(\frac{1}{r}\right)^{3/2}\left(4 \pi - \frac{157}{12} q + \eta \left[4 \pi c_0 + 2 c_{3} + q \left(2 c_{3.1}- \frac{157}{12}c_0\right) +\chi\left(-\frac{47}{4} +2c_{3.2}\right) \right]\right) \nonumber \\ &+& \left(\frac{1}{r}\right)^{2}\left(\frac{34103}{18144}+ \frac{81}{16}q^2 +\eta\left[\frac{34103}{18144}c_0 + \frac{81}{16}q^2 c_0  - \frac{3715}{1008}c_2 + \frac{7}{3} c_4 + \left(\frac{123}{8}+\frac{7}{3}\,c_{4.1}\right)q\chi\right] \right)\nonumber \\ &+&\left(\frac{1}{r}\right)^{5/2}\Bigg\{-\frac{4159}{672} \pi - \frac{1805}{84} q + \eta \bigg[ -\frac{4159}{672} \pi c_0 + \frac{20}{3}\pi\,c_2 -\frac{743}{168}c_3+\frac{8}{3}c_5+ q\left(- \frac{1805}{84}c_0 -\frac{761}{36}c_2 -\frac{743}{168}c_{3.1} + \frac{8}{3}c_{5.1}\right) \nonumber\\&+&\chi\left(\frac{3187}{672}-\frac{743}{168}c_{3.2} + \frac{8}{3}c_{5.2}\right) \bigg]\Bigg\} \Bigg\},\nonumber\\
\label{odot}
\end{eqnarray}

\noindent where, as before, \(\hat{s}= \eta \chi\). To relate the kludge and PN formalisms we propose the following coordinate transformation

\begin{eqnarray}
r &=& \frac{R}{M} \Bigg\{ 1 + b_2 \left(\frac{M}{R}\right) + \left(b_{3} +  b_{3.1}\,q \right) \left(\frac{M}{R}\right)^{3/2} + b_4 \left(\frac{M}{R}\right)^{2} + \left(b_{5} + b_{5.1}\,q \right)\left(\frac{M}{R}\right)^{5/2}\nonumber\\  &+& \eta \bigg[ b_0 +b_{2.1}\left(\frac{M}{R}\right) + (b_{3.2} + b_{3.3}\,q+  \chi\,b_{3.4}) \left(\frac{M}{R}\right)^{3/2} + \left( b_{4.1}+ b_{4.2} \,q\,\chi\right)\left(\frac{M}{R}\right)^{2} \nonumber\\ &+&  (b_{5.2} + b_{5.3}\,q+  b_{5.4}\,\chi) \left(\frac{M}{R}\right)^{5/2} \bigg]\Bigg\}, \label{coord}
\end{eqnarray}

\noindent where \(R\) stands for the PN semi-major axis. 

For consistency with the analysis outlined above, we will now use PN expressions for the frequency and its first time derivative which include spin-spin and spin-orbit corrections at 2.5PN order~\cite{blanchet,buoII,buoerr1,buoerr2}

\begin{align}\label{om2}
\Omega_{{\rm PN}}^2&=\frac{m}{r^3}\,\left\{ 1
 +\frac{m}{R}\left(-3+\eta\right)+\left(\frac{m}{R}\right)^2\left(6+\frac{41}{4}\eta-\frac{3\eta}{2}\left(\boldsymbol{\chi_1\cdot \chi_2} - 3\boldsymbol{\hat{L} \cdot \chi_1 \hat{L} \cdot \chi_2} \right)\right)
 \right.\nonumber\\&\qquad+\frac{1}{m^2}\left(\frac{m}{R}\right)^{3/2} \left[-5S_{\hat L}
 -3\frac{\delta
 m}{m}\Sigma_{\hat L}\right]\nonumber\\&\left.\qquad+\frac{1}{m^2}\left(\frac{m}{R}\right)^{5/2}\left[
 \left(\frac{39}{2}-\frac{23}{2}\eta\right)S_{\hat L}
 +\left(\frac{21}{2}-\frac{11}{2}\eta\right)\frac{\delta
 m}{m}\Sigma_{\hat L}\right]+ \mathcal{O}\left(\frac{1}{c^6}\right)\right\}\,,
\end{align}

\noindent where \(m = M + \mu \), \( \boldsymbol{\hat L}\) is a unit vector directed along the orbital momentum, \(\boldsymbol{\chi=\chi_1= S_1}/\mu^2\), \(\boldsymbol{q=\chi_2= S_2}/M^2\),  \(\delta m= \mu -M\), and we have used the spin variables

\begin{eqnarray}
\boldsymbol{S}&=&\boldsymbol{S_1 + S_2},\\
\label{assspins}
\boldsymbol{\Sigma}&=&m\left(\frac{\boldsymbol{S_2}}{M}- \frac{\boldsymbol{S_1}}{\mu}\right),
\label{sigspins}
\end{eqnarray}

\noindent  so that  \(S_{\hat L} = \boldsymbol{S\cdot \hat L}\), \(\Sigma_{\hat L} = \boldsymbol{\Sigma\cdot \hat L}\). 

The first order time derivative of the orbital frequency, including spin-orbit effects at 1.5PN and 2.5PN beyond the dominant approximation, is given by \cite{blanchet,buoII,buoerr1,buoerr2}

\begin{align}\label{omdotpn}
\frac{\dot{\Omega_{\rm PN}}}{\Omega_{\rm PN}^2}=&\frac{96}{5}\,\eta\,x^{5/2}\left\{ 1 +x\left(-\frac{743}{336}-\frac{11}{4}\eta\right)+4\pi x^{3/2}
 \right.\nonumber\\&\qquad+x^2\left(\frac{34103}{18144}+ \frac{81}{16}q^2+\sigma+\eta\left(\frac{13661}{2016} + \zeta q^{2}\right)\right) +\pi x^{5/2}\left(-\frac{4159}{672}-\frac{189}{8}\eta\right)
 \nonumber\\&\qquad+\frac{x^{3/2}}{m^2}\left[-\frac{47}{3}S^\text{c}_{\hat L}
 -\frac{25}{4}\frac{\delta
 m}{m}\Sigma^\text{c}_{\hat L}\right]\nonumber\\&\left.\qquad
 +\frac{x^{5/2}}{m^2}\left[
 \left(-\frac{5861}{144}+\frac{1001}{12}\eta\right)S^\text{c}_{\hat L}
 +\left(-\frac{809}{84}+\frac{281}{8}\eta\right)\frac{\delta
 m}{m}\Sigma^\text{c}_{\hat L}\right]\right\}\,,
\end{align}

\noindent where \(x=\left(m\Omega_{{\rm PN}}\right)^{2/3}\), and the secularly constant spin variables \(\boldsymbol{S^\text{c},\, \Sigma^\text{c}}\) are given by  \(\mathbf{S}^\text{c}=\mathbf{S}^\text{c}_1+\mathbf{S}^\text{c}_2\) and \(\mathbf{\Sigma}^\text{c}=m\left(\frac{1}{M}\mathbf{S}^\text{c}_2-\frac{1}{\mu}\mathbf{S}^\text{c}_1\right)\), with  \(S^\text{c}_{\hat L} = \boldsymbol{S^\text{c}\cdot \hat L}\), \(\Sigma^\text{c}_{\hat L} = \boldsymbol{\Sigma^\text{c}\cdot \hat L}\). The expressions for \(S^\text{c}_1, S^\text{c}_2\) are taken from Eqs. (7.2a) and (7.2b) of \cite{buoII}, under the restriction that the equations of motion used to build the kludge waveform model are valid when the spin of the inspiralling CO is perpendicular to the equatorial plane, and parallel to the momentum of the central object, i.e., for spinning circular-equatorial, non-precessing binaries.  Note that the constant \(\zeta\) in Eq.~\eqref{omdotpn} has been included to guarantee that the PN framework and the perturbative approach coincide in the test mass particle limit \(\eta \rightarrow 0\). The spin--spin parameter \(\sigma\) is given by 

\begin{eqnarray}
\sigma&=& \frac{\eta}{48}\left(-247 \boldsymbol{\chi_1\cdot \chi_2} +721\boldsymbol{\hat{L} \cdot \chi_1 \hat{L} \cdot \chi_2}\right).
\end{eqnarray}

In order to facilitate the comparison between kludge and PN results, we need to re-write expressions \eqref{om2}, \eqref{omdotpn}  in a way that takes the small mass--ratio limit by writing \(m = M (1+ \eta)\). Finally, to obtain the conservative corrections in Eq.~\eqref{omCCone}, we substitute relation \eqref{coord} into Eqs.~\eqref{omCCone} and \eqref{odot} and solve simultaneously for all the coefficients by comparing the kludge expressions for \(\Omega\) and \(\dot{\Omega}\) with their PN counterparts, i.e., \eqref{om2} and \eqref{omdotpn}, respectively.  We find that  that the non--vanishing coefficients are given by

\begin{eqnarray}
b_0 =- \frac{1}{4}, \qquad b_2 = 1, \qquad b_{2.1} = \frac{845}{448}, \qquad b_{3.2}= -\frac{9}{5}\pi, \quad   b_{3.3}= -\frac{91}{240},\nonumber\\
b_{4.1}=  -\frac{2 065 193}{677 376},  \qquad b_{5.2}=-\frac{34073}{13440} \pi, \quad b_{5.3}= \frac{1088707}{60480}, \quad b_{5.4} = \frac{31}{6}\nonumber\\
c_0 = \frac{1}{8}, \qquad c_2 = \frac{1975}{896}, \quad  c_{3}= -\frac{27}{10}\pi,  \quad c_{3.1}= -\frac{191}{160}, \qquad \zeta = - \frac{243}{32}, \nonumber \\ c_{4} = \frac{1 152 343}{451 584}, \quad c_5 = -\frac{46169}{8960}\pi,  \qquad  c_{5.1} = \frac{433703}{20160}, \quad c_{5.2}=\frac{25}{4}.
\label{10}
\end{eqnarray}

\noindent These are all the ingredients we require to explore the importance of including higher-order spin effects in the dynamics of compact binaries.  To recap, the IMRI kludge waveform model we have developed is based on the equations of motion derived by Saijo et al \cite{maeda}, which describe the dynamic evolution of spinning compact binaries. We generate the inspiral trajectory by evolving the geodesic parameters,  see Eq.~\eqref{6}, using the most accurate expression currently available for the flux of angular momentum, which includes perturbative corrections for spin-spin and spin-orbit couplings, and higher-order Teukolsky fits for the inspiral of spinless particles. We have derived an improved expression for the azimuthal frequency which includes small body spin effects,  and conservative self-force correction for higher-order spin effects. This asymptotic observable takes the following form when we implement all these corrections,  see Eq.~\eqref{7.1}, 

\begin{eqnarray}
\Omega= &&\frac{1}{r^{3/2} + q }\left(1-\frac{3}{2}\eta\chi\frac{ \sqrt{r}-q}{r^2 + q \sqrt{r}}\right)\Bigg\{1 +  \nonumber\\&&\eta\bigg[ \frac{1}{8}+ \frac{1975}{896} \left(\frac{1}{r}\right)- \left(\frac{27}{10}\pi+\frac{191}{160}q\right)\left(\frac{1}{r}\right)^{3/2} + \frac{1 152 343}{451 584}\left(\frac{1}{r}\right)^{2}  \nonumber\\ &+ & 
\left(-\frac{46169}{8960}\pi +\frac{433703}{20160}q + \frac{25}{4}\chi\right)\left(\frac{1}{r}\right)^{5/2}\bigg]\Bigg\}.
\label{omegacc}
\end{eqnarray}

\noindent This expression includes higher-order spin effects up to 2.5PN order. This is the highest order currently available for perturbative calculations that include spin-spin and spin-orbit corrections. Hence, for consistency in this analysis, we used PN calculations at the same order to amend the azimuthal frequency.  

In the following sections we will explore the influence of higher-order spin effects on source detection and parameter estimation. We will accomplish this using two different approaches: 

(i) In Section~\ref{s5}, we estimate the accuracy with which the spin of the inspiralling CO can be measured, and how the accuracy of parameter measurement depends on the spin of the small/big object. We do this using astrophysically relevant IMRI systems and a variety of spin configurations for the components of the binary. 

(ii) In Section~\ref{s6}, we assess the importance of the conservative self-force corrections for signal detection and parameter estimation by turning these parameters on and off  in the model.

\section{Waveform model}
\label{s3}

Following the numerical kludge waveform model developed in \cite{kludge}, and its subsequent extension in \cite{smallbody}, we will construct the numerical kludge waveforms combining flat-spacetime wave equations with the true inspiral trajectory developed in Section~\ref{s2}.   To apply the wave generation formula to the inspiral trajectory, we first identify the Boyer-Linquist coordinates of the source with flat-space polar coordinates. This scheme is inconsistent, but it is able to reproduce true gravitational waveforms in a large part of parameter space \cite{kludge}. 

The small perturbation \(h_{i j}(t)\) can be expressed in the transverse--traceless (TT) gauge as follows \cite{mtw} 

\begin{equation}
h_{i j}(t) = \frac{2}{D}\left(P_{ik}P_{jl}- \frac{1}{2}P_{ij}P_{kl}\right)\ddot  I^{kl}, \quad{\rm and} \quad P_{ij} = \eta_{ij} - \hat n_{i}\hat n_{j},
\label{13}
\end{equation}

\noindent where \(\eta_{ij}\) stands for the flat Minkowski metric, \(D\) is the distance to the source,  \(\hat n_{i}\) is a unit vector in the direction of propagation, and \(I^{kl}\) is the inertia tensor \cite{mtw}. 

With regard to the LISA's response function used in these studies, we notice that since most of the SNR of the IMRIs considered in this work accumulates at frequencies \(f\lesssim 10\)mHz, it is adequate to use the low-frequency approximation to the detector response which was introduced in \cite{cutlerold}, and which is summarized in Section IIIA of \cite{smallbody}.

\section{Noise induced parameter errors}
\label{s4}

To estimate the accuracies with which GW observations will be able to determine the system parameters,  \(\theta^i\), we use the Fisher Matrix formalism in the limit of high SNR. Given the expected sensitivity of LISA, this is appropriate for the systems we will consider in Section~\ref{s5}.  For large SNR, the covariance of the posterior probability distribution, \((\Gamma^{-1})^{ij}\), gives the expectation value of the errors \(\Delta \theta^i\)

\begin{equation}
\label{29}
\left< {\Delta \theta^i} {\Delta \theta^j}
 \right>  = (\Gamma^{-1})^{ij} + {\cal O}({\rm SNR})^{-1} ,
\end{equation}

\noindent where the Fisher Information Matrix \(\Gamma^{ij}\) is given by 

\begin{equation}
\label{27}
\Gamma_{ij} \equiv \bigg( \frac{\partial {\bf h}}{ \partial \theta^i}\, \bigg| \,
\frac{\partial {\bf h}}{ \partial \theta^j }\bigg)_{|\theta=\hat{\theta}},
\end{equation}

\noindent where \(\hat{\theta}\) is the true value of the parameters of the source. For white noise, i.e., \(S_n(f)= \textrm{const.}\), Parseval's theorem allows us to write the inner product of two given signals, \(\left( {\bf p} \,|\, {\bf q} \right)\), as  \(2 S_n^{-1}\sum_{\alpha}\int_{-\infty}^{\infty} \, p_\alpha(t) q_\alpha(t) \mathrm{d}t\) \cite{cutler}. Hence, following Barack \& Cutler \cite{cutler},  we can define the ``noise--weighted'' waveform as follows

\begin{equation}\label{30}
\hat h_{\alpha}(t) \equiv   \frac{h_{\alpha}(t)}{\sqrt{S_h\bigl(f(t)\bigr)}}, \qquad f(t) = \frac{1}{\pi}\frac{\mathrm{d}\phi}{\mathrm{d}t}.
\end{equation}

\noindent This convention enable us to find the following approximate expression for the Fisher matrix 

\begin{equation}\label{31}
\Gamma_{ab} = 2\sum_{\alpha}\int_0^T{\partial_a \hat h_{\alpha}(t) \partial_b \hat h_{\alpha}(t) \mathrm{d}t} \, .
\end{equation}

\noindent Using the same prescription presented in Section V(a) of Barack and Cutler~\cite{cutler}, the total LISA noise function, \(S_h\bigl(f\bigr)\),  has three components: instrumental noise, confusion noise from short--period galactic binaries, and confusion noise from extragalactic binaries.

\section{Parameter estimation error results}
\label{s5}

We shall now make use of the Fisher Matrix formalism introduced above to estimate the accuracy with which LISA will be able to determine the spin of IMBHs that inspiral into SMBHs. The IMRI kludge waveform model used in these studies includes all the higher-order spin effects described in Section~\ref{s2}. We  present the parameter error estimates for two astrophysically relevant types of IMRI, assuming three different spin configurations: rapid spins, moderate spins, and low spins. 

It is worth pointing out that the parameter estimation results we present in the following Sections were obtained from Fisher Matrices which were nicely convergent over several orders of magnitude in the offset used to compute the numerical derivatives (see Eq.~\eqref{27}). We also checked that the inverses of these matrices had good convergent behavior.  

We will estimate the noise-induced errors by fixing the values of the intrinsic parameters of the source and running a Monte Carlo simulation over possible values of the extrinsic parameters. To estimate the SNR associated with each event we use the relation

\begin{equation}\label{35}
{\rm SNR}^2= 2\sum_{\alpha=I,II}\int_{t_{\rm init}}^{t_{\rm LSO}}
\hat h_{\alpha}^2(t)\mathrm{d}t,
\end{equation}

\noindent  and assume that the distance to the source is $D=1$Gpc. We use these Monte Carlo results to obtain the SNR distribution of each type of source. This information will allow us to find a `typical' SNR value to renormalise the parameter estimation results. We describe this method in detail in the following Section.

\subsection{Determination of typical SNRs}
\label{snrsec}

To determine the appropriate SNR at which we quote the parameter estimation errors for each type of source, we have carried out a Monte Carlo simulation over values of the extrinsic parameters of the source, while fixing the value of the intrinsic parameters. The events we consider were uniformly distributed in comoving volume out to a redshift of $z=1$, and were assumed to be observed over a one year time window at the detector.  Under these assumptions, we computed the SNR for each event, then looked at the SNR distribution of the detected events and chose the ``typical'' reference SNR as the median of the SNR distribution.  The events considered in this paper could be detected by LISA at redshifts $z>1$. However, parameter determination for these events would involve the use of models that describe the formation channels of IMBHs and the inspiral rates --information which is currently uncertain. Hence, we shall consider events which take place only at redshifts $z \lesssim 1$. 

To ensure that the events are observed by LISA over one year, we choose the initial radius of inspiral, \(p_0\), such that at the end of the observation time window the inspiralling IMBH has reached the innermost stable circular orbit. We use this information as input data in the Monte Carlo simulations. We also fix the redshifted masses of the sources to the values of the two IMRI systems we will consider later, i.e., $5\times 10^3M_\odot+10^6M_{\odot}$ and $10^4M_\odot+10^6M_{\odot}$.

\begin{figure*}[ht]
\centerline{
\includegraphics[height=0.33\textwidth,  clip]{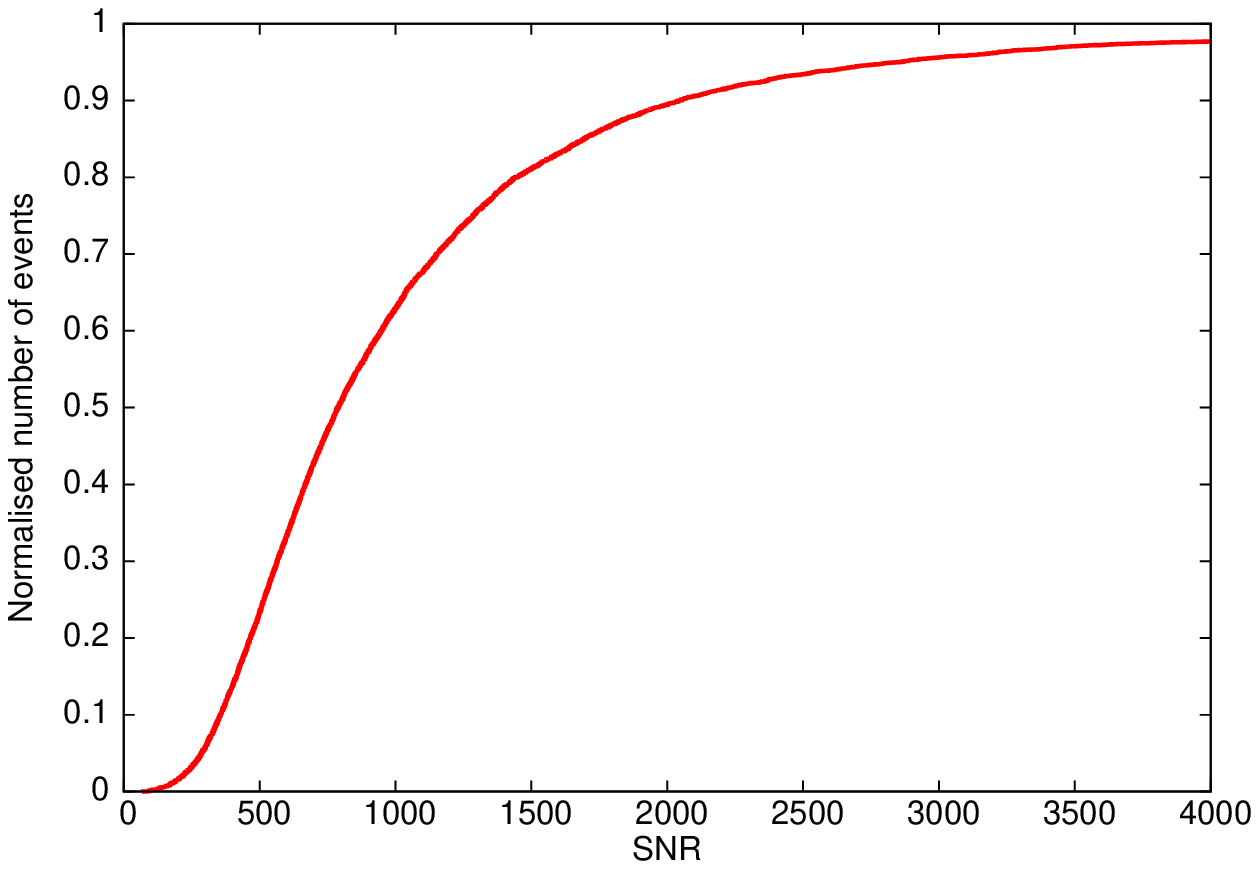}
}
\centerline{
\includegraphics[height=0.33\textwidth,  clip]{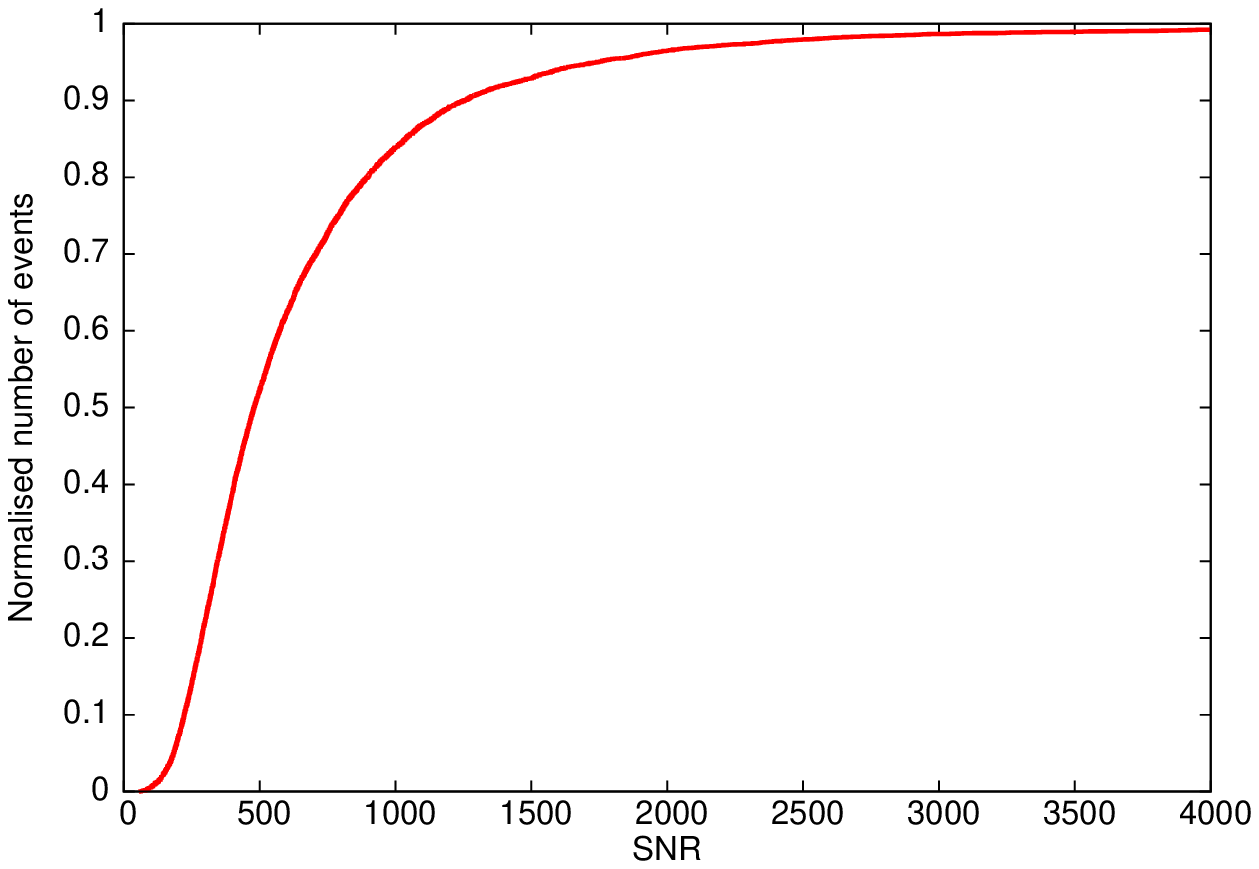}
\includegraphics[height=0.33\textwidth,  clip]{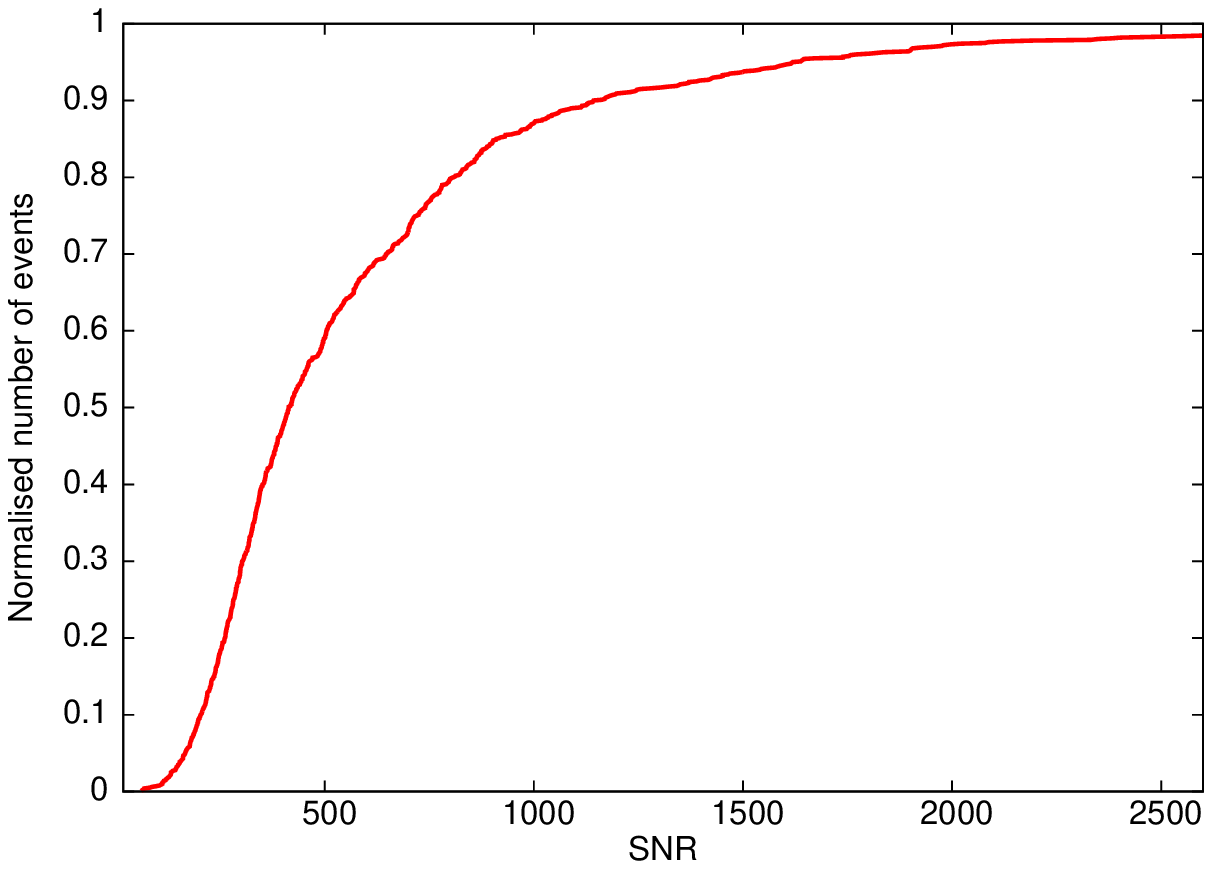}
}
\caption{Normalised cumulative distribution function for the signal--to--noise ratio of a cosmological population of binary systems with redshifted masses  $\mu=5\times10^3M_{\odot},\, M=10^6M_{\odot}$. The inspiralling IMBHs were taken to have spin parameter \(\chi =0.9\), while the spin parameter of the central SMBH is  \(q=0.9\) (top panel), \(q=0.3\) (bottom left) and \(q=0.1\) (bottom right). }
\label{smoothdis}
\end{figure*}

Figure~\ref{smoothdis} shows the normalized cumulative distribution functions  for the SNRs computed in this way, for binary systems with central SMBHs of redshifted mass  $10^6M_{\odot}$ and spin parameters $q=0.9,\,0.3,\,0.1$. The inspiralling IMBHs have specific spin parameters  \(\chi =0.9, \,0.3,\,0.1\), and redshifted mass $\mu=5\times10^3M_{\odot}$.  We find that the median SNR for sources with spin parameters  $(\chi=0.3,\, 0.1)$ is approximately the same as that obtained for  \( \chi =0.9 \),  so we only show the SNR distribution for this representative case for each type of source. This is true for all three spins of the central SMBH.

\begin{figure*}[ht]
\centerline{
  \includegraphics[height=0.33\textwidth,angle=0,  clip]{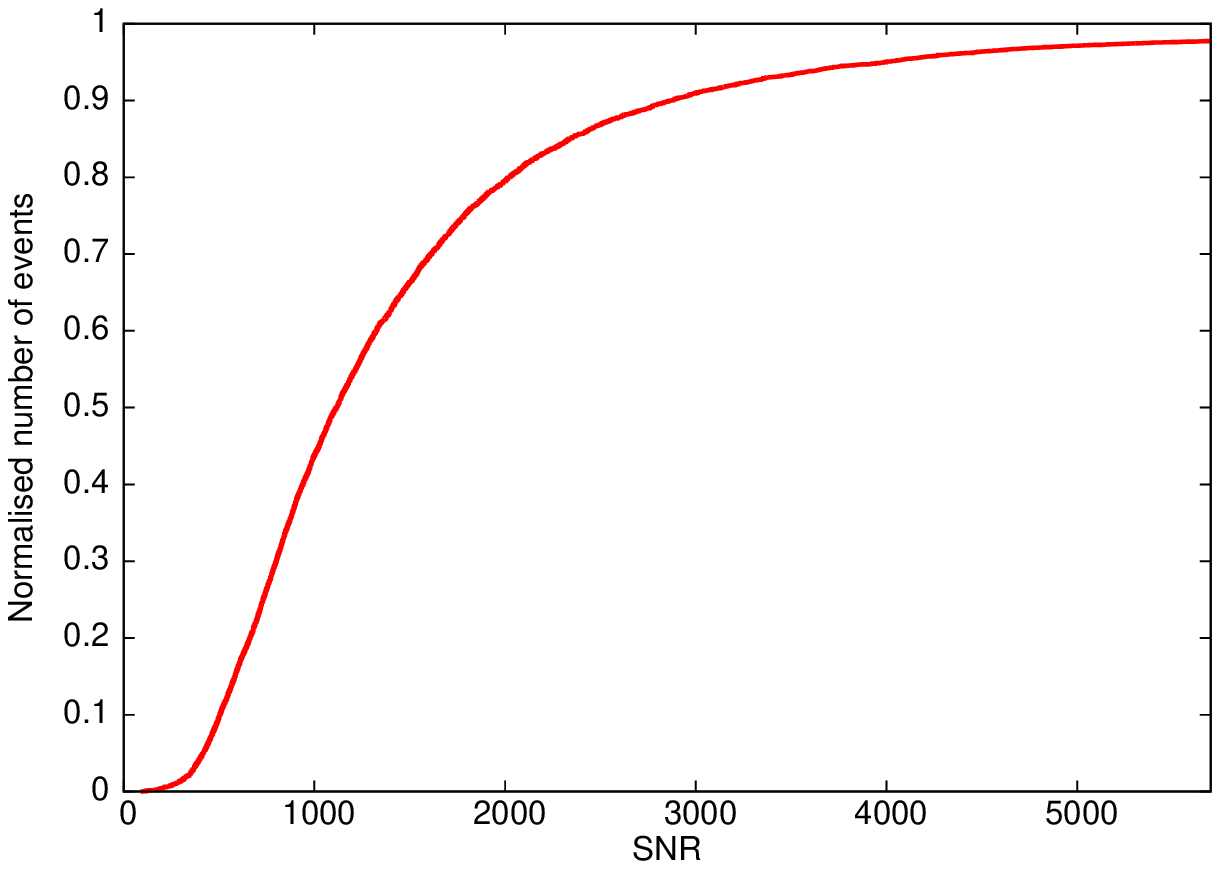}
}
\centerline{
\includegraphics[height=0.33\textwidth,  clip]{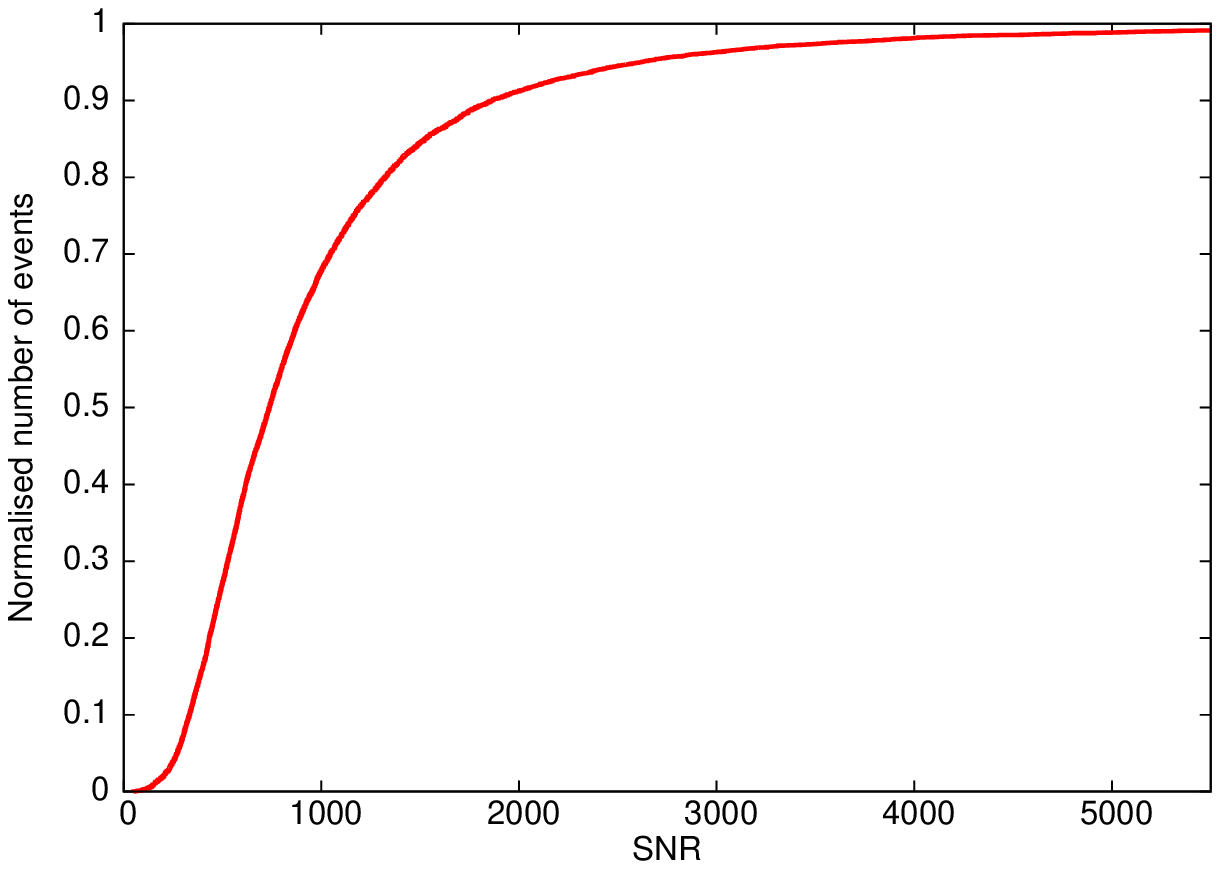}
\includegraphics[height=0.33\textwidth,  clip]{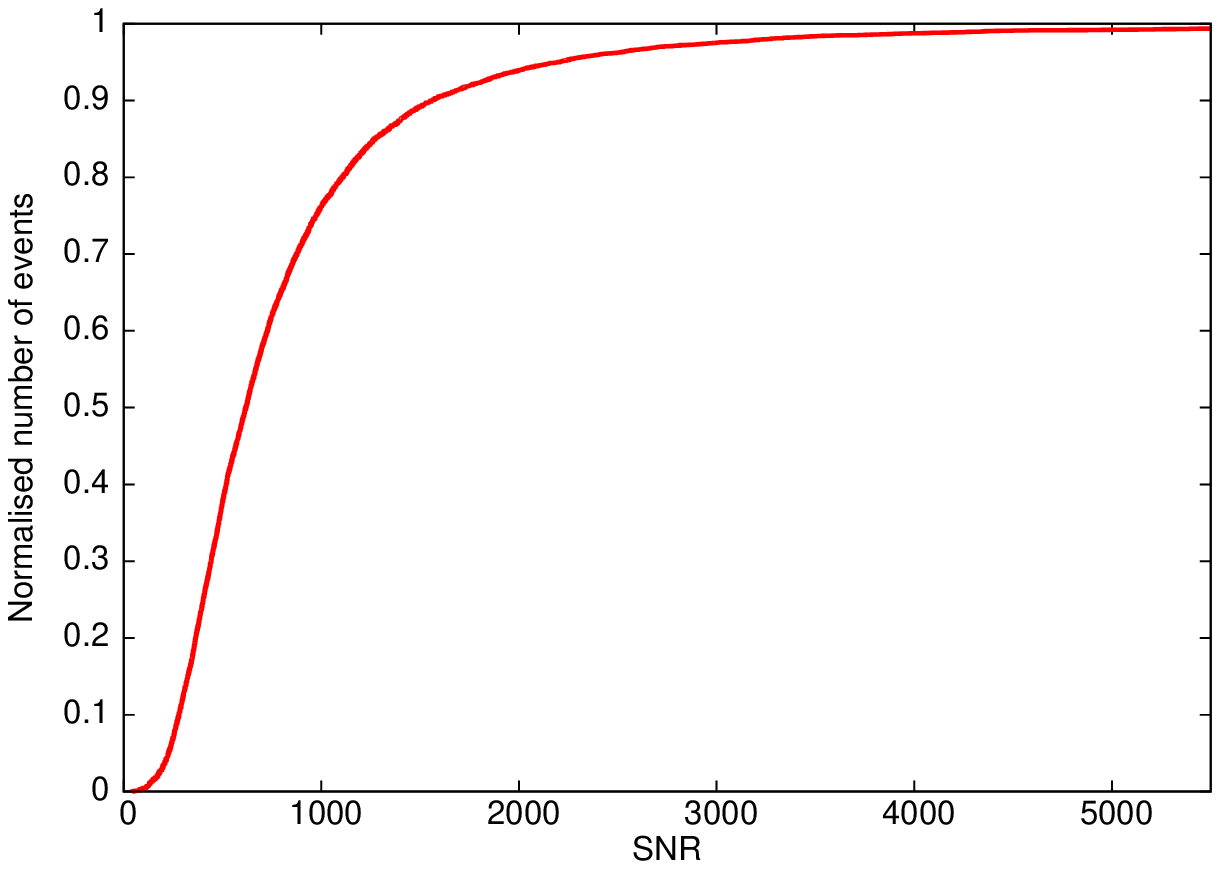}
}
\caption{Normalized cumulative distribution function for the signal--to--noise ratio of a cosmological population of binary systems with BHs of redshifted mass  $\mu=10^4M_{\odot}, \, M=10^6M_{\odot}$. The panels show results for three different combinations of the spin magnitudes of the central and inspiralling BHs, \((q,\chi)\), namely: top panel \((0.9, 0.9)\); bottom-left panel \((0.3,0.9)\); bottom-right panel \((0.1, 0.9)\).}
\label{snrs}
\end{figure*}

In Figure~\ref{snrs}  we present the normalized cumulative distribution function for  binary systems with redshifted masses $10^4M_{\odot}+10^6M_{\odot}$, for the same spin configurations used in Figure~\ref{smoothdis}.

The results presented in Figure~\ref{smoothdis} show that the ``typical'' SNR of sources with redshifted masses \( 5\times10^3M_{\odot} + 10^6M_{\odot}\), and spin parameters \((q,\, \chi)= (0.9,\{0.9,\,0.3,\,0.1\})\);  \((q,\, \chi)= (0.3,\{0.9,\,0.3,\,0.1\})\); and \((q,\, \chi)= (0.1,\{0.9,\,0.3,\,0.1\})\) are \(1000,\, 500, \,500\), respectively. Similarly, Figure~\ref{snrs} suggests that the median SNRs of the binaries with component masses \( 10^4M_{\odot} + 10^6M_{\odot}\) and spin parameters \((q,\, \chi)= (0.9,\{0.9,\,0.3,\,0.1\})\);  \((q,\, \chi)= (0.3,\{0.9,\,0.3,\,0.1\})\) and \((q,\, \chi)= (0.1,\{0.9,\,0.3,\,0.1\})\) are \(1000,\, 750, \,750\), respectively.  We will use these reference values to renormalise the parameter estimation errors presented in Tables~\ref{tabFMErrMBH}-\ref{tabFMErrIMBH01} so that they represent the typical parameter estimation errors for a cosmological population of sources at $z \lesssim 1$. For the top panel of Figure~\ref{smoothdis}, SNR\(=1000\) corresponds to the 60-th percentile of the distribution rather the median, but we have chosen this reference value as it is a convenient reference point which is easy to readjust to other SNRs. The results in Figures~\ref{smoothdis} and~\ref{snrs} could also be used to renormalise the results to a different SNR reference value if required.

\subsection{Parameter estimation results}

In this Section we present parameter estimation results for the 11 parameters ---see Table~\ref{tableparams}--- that characterise these systems.  The results obtained in \cite{smallbody} suggested that for IMRIs with mass ratio \(\eta\gtrsim10^{-3}\), LISA observations might allow us to measure the spin of an inspiralling IMBH to a fractional accuracy of \(\sim 10\%\) if observed with an SNR of 1000, provided both components of the binary were rapidly rotating, i.e, for \(q=\chi=0.9\). In this section, we extend that analysis and present results that suggest that GW observations could still provide information about the spins of IMBHs even if the spin of the inspiralling IMBH is not high.

\begin{table}[thb]
\centerline{$\begin{array}{c|l}\hline\hline
\ln \mu  &\text {mass of inspiralling object}   \\
\ln M    &\text {mass of central SMBH}\\
q        &\text{magnitude of (specific) spin angular momentum of SMBH} \\
\chi     &\text{magnitude of (specific) spin angular momentum of inspiralling object} \\
p_0      &\text{Initial radius of inspiralling object's orbit} \\
\phi_0   &\text{Initial phase of  inspiralling object's orbit}      \\
\theta_S &\text{source sky colatitude in an ecliptic--based system }  \\
\phi_S   &\text{source sky azimuth in an ecliptic--based system}  \\
\theta_K  &\text{direction  of SMBH spin (colatitude)}  \\
\phi_K   &\text{direction of SMBH spin (azimuth)}  \\
\ln D    &\text{distance to  source}\\
\hline\hline
\end{array}$}
\caption{\protect\footnotesize
This table describes the meaning of the parameters used in our model. The angles ($\theta_S$,\,$\phi_S$) and ($\theta_K$,\,$\phi_K$) are defined in a fixed ecliptic--based coordinate system.}
\label{tableparams}
\end{table}

In Table~\ref{tabFMErrMBH}, we can see that for IMRIs with mass components  \( 5\times10^3M_{\odot}+10^6M_{\odot}\), GW observations could determine the spin of IMBHs to a fractional accuracy of \(\sim 10\%\) at a fixed SNR=\(1000\) when the central SMBH is rapidly rotating $(q=0.9)$ and the spin of the IMBH is moderate to high \((\chi=0.9,\,0.3)\).  This is the first relevant result of this paper, and constitutes an important extension to the analysis originally presented in \cite{smallbody}, which did not include parameter estimation results for binaries with spin parameters \((q=0.9,\, \chi =0.3\)). It is worth pointing out that this new finding does not crucially depend on the inclusion of higher-order spin effects in the waveform template. We reach this conclusion by comparing the results of Table V in \cite{smallbody} with those of Table~\ref{tabFMErrMBH} in this paper, which suggest that  the inclusion of higher-order spin effects will have a minor impact on the accuracy with which GW observations may determine the source parameters of \( 5\times10^3M_{\odot}+10^6M_{\odot}\) IMRIs. We shall see below that  the inclusion of higher-order spin corrections in search templates do become important for parameter estimation of \(10^4M_{\odot}+10^6M_{\odot}\) IMRIs.

Table~\ref{tabFMErrMBH} also shows that LISA observations could also constrain the spin of IMBHs with \(\chi\sim 0.1\), but only to an accuracy  \(\sim25\%\).  Tables~\ref{tabFMErrMMBH}, \ref{tabFMErrMMBHs} show the corresponding results for cases in which the binary has the same component masses as in Table~\ref{tabFMErrMBH}, but the central SMBH has spin parameter \(q=0.3,\,0.1\), respectively. These Tables confirm that we are unlikely to accurately measure the spin of an IMBH that inspirals into an SMBH with moderate or low spin. 

\begin{table}[thb]
\begin{tabular}{|c|c|c|c|c|c|c|c|c|c|c|c|c|}
\hline\multicolumn{2}{|c|}{}&\multicolumn{11}{c|}{Distribution of \(\log_{10}(\Delta X)\) in error, \(\Delta X\), for parameter \(X=\)}\\\cline{3-13}
\multicolumn{2}{|c|}{Model}&$\ln(m)$&$\ln(M)$&$q$&$\chi$&$p_0$&$\phi_0$&$\theta_S$&$\phi_S$&$\theta_K$&$\phi_K$&$\ln(D)$\\\hline
&Mean                      &-3.14&   -2.97&  -4.37&  -1.08&  -1.67&  -1.55&   -2.18&  -2.13&   -1.85&  -1.75&  -1.92\\\cline{2-13}
$q=0.9$&St. Dev.    &0.076& 0.94&    0.078&0.071&  0.085& 0.438&  0.467& 0.345& 0.434& 0.530&  0.385\\\cline{2-13}
&L. Qt.                       &-3.15&  -2.96&   -4.39&  -1.10&  -1.65&  -1.90&   -2.43&  -2.35&  -2.13&  -2.10&   -2.11\\\cline{2-13}
\(\chi=0.9\)&Med.     &-3.13&  -2.95&   -4.37&  -1.06&  -1.65&  -1.71&   -2.02&  -2.01&  -1.89&  -1.80&   -1.95\\\cline{2-13}
&U. Qt.                        &-3.06&  -2.82&  -4.28&  -1.01&  -1.53&  -1.46&   -1.81&  -1.79&  -1.62&  -1.58&   -1.73\\\hline

&Mean                       &-3.12&  -2.94&   -4.38&   -1.05&  -1.54&   -1.45&   -2.06&   -2.03&   -1.84&   -1.71&   -1.91\\\cline{2-13}
$q=0.9$&St. Dev.    &0.059&  0.073&  0.053&  0.048& 0.074&  0.439&  0.427&  0.455&  0.381&  0.432&  0.297\\\cline{2-13}
&L. Qt.                        &-3.14& -2.95&   -4.39&   -1.06&   -1.63&   -1.89&   -2.34&   -2.24&   -2.14&   -2.03&   -2.10\\\cline{2-13}
\(\chi=0.3\)&Med.      &-3.13& -2.93&   -4.38&   -1.04&   -1.55&   -1.45&   -1.98&   -1.93&   -1.86&   -1.78&  -1.93\\\cline{2-13}
&U. Qt.                        &-3.12& -2.89&   -4.37&   -1.03&   -1.52&   -1.46&   -1.76&   -1.72&   -1.63&   -1.51&   -1.74\\\hline

&Mean                        &-2.70&  -2.51&  -3.98&  -0.65&  -1.22&  -1.46&   -2.08&  -2.06&  -1.79&  -1.66&  -1.86\\\cline{2-13}
$q=0.9$&St. Dev.     &0.124&  0.141&0.151& 0.163&  0.221& 0.378& 0.449& 0.347& 0.391& 0.457& 0.495\\\cline{2-13}
&L. Qt.                         &-2.85&  -2.65&  -4.16&  -0.82&  -1.37&   -1.88&  -2.43&  -2.35&  -2.11&  -1.96&  -2.06\\\cline{2-13}
\(\chi=0.1\)&Med.      &-2.64&  -2.45&  -3.92&  -0.63&  -1.22&   -1.52&  -2.03&   -2.01&  -1.81&  -1.73&  -1.89\\\cline{2-13}
&U. Qt.                        &-2.48&  -2.32&  -3.79&  -0.46&  -1.12&   -1.37&  -1.77&   -1.74&  -1.63&  -1.43&  -1.63\\\hline
\end{tabular}
\caption{ Summary of results of the Monte Carlo simulation of Fisher Matrix errors for spinning BH systems with masses $\mu=5\times10^3M_{\odot}$, \(M=10^6M_{\odot}\) and for a central SMBH with spin parameter \(q=0.9\). We show the mean, standard deviation, median and quartiles of the distribution of the logarithm to base ten of the error in each parameter. Results are given for the kludge model with conservative corrections to 2.5PN order and quoted at a fixed SNR$=1000$.}
\label{tabFMErrMBH}
\end{table}

\begin{table}[thb]
\begin{tabular}{|c|c|c|c|c|c|c|c|c|c|c|c|c|}
\hline\multicolumn{2}{|c|}{}&\multicolumn{11}{c|}{Distribution of \(\log_{10}(\Delta X)\) in error, \(\Delta X\), for parameter \(X=\)}\\\cline{3-13}
\multicolumn{2}{|c|}{Model}&$\ln(m)$&$\ln(M)$&$q$&$\chi$&$p_0$&$\phi_0$&$\theta_S$&$\phi_S$&$\theta_K$&$\phi_K$&$\ln(D)$\\\hline
&Mean                       &-3.10&  -2.90&   -2.70&   -0.29&  -1.59&   -1.52&   -1.95&   -1.87&   -1.80&   -1.67&   -1.80\\\cline{2-13}
$q=0.3$&St. Dev.    &0.080&  0.084&  0.075&  0.064& 0.081&  0.375&  0.413&  0.463&  0.348&  0.402&  0.250\\\cline{2-13}
&L. Qt.                        &-3.12& -2.95&   -2.76&   -0.34&   -1.63&   -1.61&   -2.18&   -2.07&   -2.01&   -1.93&   -1.96\\\cline{2-13}
\(\chi=0.9\)&Med.      &-3.09& -2.92&   -2.73&   -0.30&   -1.59&   -1.51&   -1.86&   -1.78&   -1.79&   -1.66&  -1.83\\\cline{2-13}
&U. Qt.                        &-3.00& -2.82&   -2.64&   -0.23&   -1.51&   -1.43&   -1.64&   -1.58&   -1.58&   -1.42&   -1.66\\\hline

&Mean                       &-3.10&  -2.91&   -2.66&   -0.28&  -1.59&   -1.53&   -1.92&   -1.85&   -1.77&   -1.63&   -1.78\\\cline{2-13}
$q=0.3$&St. Dev.    &0.073&  0.098&  0.057&  0.062& 0.087&  0.291&  0.397&  0.441&  0.344&  0.390&  0.247\\\cline{2-13}
&L. Qt.                        &-3.13& -2.94&   -2.73&   -0.32&   -1.58&   -1.62&   -2.14&   -2.07&   -1.98&   -1.88&   -1.93\\\cline{2-13}
\(\chi=0.3\)&Med.      &-3.11& -2.90&   -2.64&   -0.28&   -1.56&   -1.52&   -1.82&   -1.78&   -1.75&   -1.62&  -1.81\\\cline{2-13}
&U. Qt.                        &-3.01& -2.85&   -2.62&   -0.24&   -1.53&   -1.42&   -1.60&   -1.57&   -1.55&   -1.39&   -1.64\\\hline

&Mean                       &-3.07&  -2.87&   -2.63&   -0.20&  -1.59&   -1.45&   -1.91&   -1.85&   -1.77&   -1.62&   -1.78\\\cline{2-13}
$q=0.3$&St. Dev.    &0.081&  0.097&  0.093&  0.088& 0.102&  0.439&  0.421&  0.434&  0.349&  0.376&  0.245\\\cline{2-13}
&L. Qt.                        &-3.11& -2.92&   -2.72&   -0.25&   -1.61&   -1.89&   -2.21&   -2.10&   -1.99&   -1.84&   -1.93\\\cline{2-13}
\(\chi=0.1\)&Med.      &-3.03& -2.84&   -2.65&   -0.21&   -1.57&   -1.45&   -1.84&   -1.78&   -1.77&   -1.63&  -1.80\\\cline{2-13}
&U. Qt.                        &-3.00& -2.79&   -2.61&   -0.17&   -1.49&   -1.46&   -1.61&   -1.56&   -1.55&   -1.37&   -1.63\\\hline
\end{tabular}
\caption{ As Table~\ref{tabFMErrMBH}, but for a central SMBH with spin parameter \(q=0.3\). Results are quoted at a fixed SNR of 500.}
\label{tabFMErrMMBH}
\end{table}

\begin{table}[thb]
\begin{tabular}{|c|c|c|c|c|c|c|c|c|c|c|c|c|}
\hline\multicolumn{2}{|c|}{}&\multicolumn{11}{c|}{Distribution of \(\log_{10}(\Delta X)\) in error, \(\Delta X\), for parameter \(X=\)}\\\cline{3-13}
\multicolumn{2}{|c|}{Model}&$\ln(m)$&$\ln(M)$&$q$&$\chi$&$p_0$&$\phi_0$&$\theta_S$&$\phi_S$&$\theta_K$&$\phi_K$&$\ln(D)$\\\hline
&Mean                      &-3.09&  -2.92&  -2.61&  -0.16&  -1.62&  -1.58&  -1.94&   -1.85&   -1.73&  -1.62&  -1.76\\\cline{2-13}
$q=0.1$&St. Dev.   &0.075& 0.072& 0.065& 0.063&  0.073& 0.380&  0.416& 0.434& 0.396& 0.443&  0.306\\\cline{2-13}
&L. Qt.                       &-3.16&  -2.99&  -2.67&  -0.20&   -1.68&  -1.80&   -2.24&  -2.15&  -1.96&  -1.86&  -1.96\\\cline{2-13}
\(\chi=0.9\)&Med.    &-3.09&  -2.92&  -2.61&  -0.16&   -1.62&   -1.61&   -1.92&  -1.87&  -1.74&  -1.63&  -1.79\\\cline{2-13}
&U. Qt.                     &-3.02&   -2.86&  -2.57&  -0.12&   -1.57&   -1.42&   -1.68&  -1.63&  -1.49&  -1.39&  -1.59\\\hline

&Mean                      &-3.09&  -2.92&   -2.61&  -0.16&  -1.62&  -1.58&   -1.94&   -1.84&   -1.71&  -1.64&   -1.76\\\cline{2-13}
$q=0.1$&St. Dev.   &0.092&  0.078&  0.075& 0.068& 0.083& 0.376&  0.409&  0.425&  0.402& 0.385&   0.350\\\cline{2-13}
&L. Qt.                       &-3.16&   -2.99&   -2.67&  -0.21&  -1.69&  -1.85&  -2.22&   -2.11&   -2.02&   -1.85&   -1.95\\\cline{2-13}
\(\chi=0.3\)&Med.    &-3.08&   -2.92&   -2.61&  -0.16&  -1.61&  -1.60&   -1.93&   -1.86&   -1.71&   -1.63&   -1.79\\\cline{2-13}
&U. Qt.                      &-3.03&   -2.85&   -2.56&  -0.12&  -1.54&  -1.49&   -1.69&   -1.62&   -1.50&   -1.45&   -1.52\\\hline

&Mean                      &-3.09&  -2.92&   -2.61&  -0.16&  -1.62&  -1.59&   -1.95&   -1.84&   -1.69&  -1.64&   -1.75\\\cline{2-13}
$q=0.1$&St. Dev.   &0.081&  0.079&  0.072& 0.079& 0.076& 0.311&  0.376&  0.493&  0.394& 0.392&   0.492\\\cline{2-13}
&L. Qt.                       &-3.16&   -2.99&   -2.67&  -0.21&  -1.69&  -1.85&  -2.22&   -2.11&   -2.02&   -1.85&   -1.95\\\cline{2-13}
\(\chi=0.1\)&Med.    &-3.10&   -2.92&   -2.62&  -0.17&  -1.61&  -1.64&   -1.95&   -1.85&   -1.72&   -1.63&   -1.79\\\cline{2-13}
&U. Qt.                      &-3.02&   -2.86&   -2.56&  -0.12&  -1.54&  -1.49&   -1.69&   -1.62&   -1.50&   -1.45&   -1.52\\\hline
\end{tabular}
\caption{ As Table~\ref{tabFMErrMBH}, but for a slowly rotating central BH with spin parameter \(q=0.1\). Results are quoted at a fixed SNR of 500. Note that for slowly rotating SMBHs, the spin magnitude of the inspiralling IMBH has no noticeable influence on the accuracy with which the source parameters can be determined. }
\label{tabFMErrMMBHs}
\end{table}

\clearpage

Table~\ref{tabFMErrIMBH} shows that for IMRIs with mass components \(10^4M_{\odot}+10^6M_{\odot}\) and whose central SMBH is rapidly rotating, GW observations may allow us to obtain an accurate measurement of the spin distribution of IMBHs. These results show that at a fixed SNR\(=1000\),  the spin of IMBHs could be determined to a fractional accuracy better than \(10\%\) for the spin combinations \((q=0.9, \{\chi=0.9,\,0.3,\,0.1\})\).

Table~\ref{tabFMErrIMBH03} shows that for IMRIs with the same component masses considered in Table~\ref{tabFMErrIMBH}, but with moderately rotating central SMBHs, LISA observations may enable us to constrain the spin of rapidly rotating IMBHs to an accuracy better than \(\sim 25\%\), at a fixed SNR=\(750\). If, on the other hand, the IMBH is either moderately or slowly rotating, we could at least be able to constrain the spin parameter to an accuracy \(\sim 30\%\), at a fixed SNR=\(750\). We also note that these results for the spin of the IMBH are a factor of \(\sim 2\) better than those shown in Table~\ref{tabFMErrMMBH}. 

Finally,  Table~\ref{tabFMErrIMBH01} shows that for IMRIs with mass components \(10^4M_{\odot}+10^6M_{\odot}\), and a slowly rotating central SMBH, LISA observations will not allow an accurate extraction of the spin of IMBHs. The best we could do is to constrain this parameter to better than \(40\%\), at a fixed SNR\(=750\). 

To recap, the analysis of these results suggests that, (i) for IMRI sources with mass ratio \(\eta\sim10^{-3}\), observed at a typical SNR\(=1000\), GW observations may allow us to measure the spin of IMBHs with rapid and moderate spins when the central SMBH is rapidly rotating to an accuracy  \(\sim10\%\);  (ii) for IMRI sources with mass ratios \(\eta\sim10^{-2}\), observed at a typical SNR\(=1000\), LISA observations will enable us to determine the spin of IMBHs whether they are rapidly, moderately or slowly rotating, as long as the central SMBH is rapidly rotating, to an accuracy  better than \(\sim10\%\);  (iii) for IMRI sources with \(\eta\sim10^{-2}\),  but with moderately rotating SMBHs, LISA observations will enable us to constrain the spin of IMBHs to an accuracy better than \(30\%\), at a fixed SNR\(=750\), even if the IMBH is slowly rotating. 

This improved analysis also confirms our results in \cite{smallbody}, namely, the determination of the spin of the inspiralling object through GW observations is mostly determined by the effect that spin couplings have on the orbital evolution. Our initial analysis in \cite{smallbody} indicated that the spin coupling can be best determined when the central SMBH has high spin.  However, in light of these new results, we conclude that while we still require the central object to rotate rapidly, the companion does not need to have high spin. Indeed, we have found that when we observe IMRIs with mass ratios \(\eta\sim10^{-2}\) with sufficient SNR, small body spin effects become significant enough on the orbital evolution of IMRIs that we can get accurate information about the spin distribution of slowly rotating IMBHs.

These results are astrophysically  meaningful, since the studies presented in~\cite{mandel} suggest that IMBHs that grow through a series of minor mergers are expected to have typical spins \(\chi\sim0.3\). Hence, GW observations will not only shed light on the existence of IMBHs, but may also provide information  about their spin distribution --an astrophysical property that is determined by their channel of formation. These findings would, in turn, be useful  to refine models of BH growth.

\begin{table}[thb]
\begin{tabular}{|c|c|c|c|c|c|c|c|c|c|c|c|c|}
\hline\multicolumn{2}{|c|}{}&\multicolumn{11}{c|}{Distribution of \(\log_{10}(\Delta X)\) in error, \(\Delta X\), for parameter \(X=\)}\\\cline{3-13}
\multicolumn{2}{|c|}{Model}&$\ln(m)$&$\ln(M)$&$q$&$\chi$&$p_0$&$\phi_0$&$\theta_S$&$\phi_S$&$\theta_K$&$\phi_K$&$\ln(D)$\\\hline
&Mean                       &-3.05&  -2.87&   -4.09&   -1.12&  -1.51&   -1.58&   -2.10&   -2.04&   -1.86&   -1.72&   -1.89\\\cline{2-13}
$q=0.9$&St. Dev.    &0.039&  0.062&  0.045&  0.052& 0.063&  0.333&  0.454&  0.470&  0.382&  0.411&  0.287\\\cline{2-13}
&L. Qt.                        &-3.06& -2.88&   -4.11&   -1.15&   -1.52&   -1.80&   -2.36&   -2.30&   -2.11&   -1.98&   -2.05\\\cline{2-13}
\(\chi=0.9\)&Med.      &-3.04& -2.87&   -4.09&   -1.11&   -1.51&   -1.63&   -1.97&   -1.93&   -1.85&   -1.77&  -1.93\\\cline{2-13}
&U. Qt.                        &-3.01& -2.85&   -4.07&   -1.09&   -1.49&   -1.48&   -1.74&   -1.72&   -1.67&   -1.55&   -1.75\\\hline

&Mean                       &-3.03&  -2.87&   -4.06&   -1.04&  -1.48&   -1.58&   -2.07&   -1.99&   -1.87&   -1.73&   -1.89\\\cline{2-13}
$q=0.9$&St. Dev.    &0.096&  0.104&  0.101&  0.097& 0.085&  0.311&  0.445&  0.494&  0.369&  0.399&  0.283\\\cline{2-13}
&L. Qt.                        &-3.05& -2.89&   -4.09&   -1.09&   -1.54&   -1.79&   -2.29&   -2.24&   -2.11&   -1.96&   -2.04\\\cline{2-13}
\(\chi=0.3\)&Med.      &-3.04& -2.88&   -4.10&   -1.09&   -1.49&   -1.60&   -1.98&   -1.91&   -1.87&   -1.71&  -1.92\\\cline{2-13}
&U. Qt.                        &-3.02& -2.85&   -4.01&   -0.98&   -1.38&   -1.45&   -1.72&   -1.69&   -1.63&   -1.48&   -1.73\\\hline

&Mean                       &-2.92&  -2.79&   -4.00&   -0.96&  -1.39&   -1.50&   -2.07&   -2.00&   -1.88&   -1.73&   -1.89\\\cline{2-13}
$q=0.9$&St. Dev.    &0.109&  0.134&  0.074&  0.059& 0.086&  0.314&  0.443&  0.501&  0.366&  0.403&  0.282\\\cline{2-13}
&L. Qt.                        &-3.02& -2.83&   -4.09&   -1.05&   -1.40&   -1.70&   -2.31&   -2.21&   -2.08&   -1.95&   -2.05\\\cline{2-13}
\(\chi=0.1\)&Med.      &-2.91& -2.76&   -4.01&   -1.01&   -1.38&   -1.53&   -1.96&   -1.89&   -1.88&   -1.71&  -1.92\\\cline{2-13}
&U. Qt.                        &-2.84& -2.71&   -3.98&   -0.99&   -1.34&   -1.41&   -1.71&   -1.67&   -1.65&   -1.49&   -1.73\\\hline
\end{tabular}
\caption{ As Table~\ref{tabFMErrMBH}, but for an inspiralling BH with mass $\mu=10^4M_{\odot}$. Results are quoted at a fixed SNR of 1000.}
\label{tabFMErrIMBH}
\end{table}

\begin{table}[thb]
\begin{tabular}{|c|c|c|c|c|c|c|c|c|c|c|c|c|}
\hline\multicolumn{2}{|c|}{}&\multicolumn{11}{c|}{Distribution of \(\log_{10}(\Delta X)\) in error, \(\Delta X\), for parameter \(X=\)}\\\cline{3-13}
\multicolumn{2}{|c|}{Model}&$\ln(m)$&$\ln(M)$&$q$&$\chi$&$p_0$&$\phi_0$&$\theta_S$&$\phi_S$&$\theta_K$&$\phi_K$&$\ln(D)$\\\hline
&Mean                       &-2.99&  -2.84&   -2.70&   -0.63&  -1.50&   -1.59&   -2.06&   -1.97&   -1.87&   -1.72&   -1.88\\\cline{2-13}
$q=0.3$&St. Dev.    &0.050&  0.054&  0.039&  0.063& 0.055&  0.313&  0.430&  0.456&  0.358&  0.392&  0.274\\\cline{2-13}
&L. Qt.                        &-3.07& -2.90&   -2.72&   -0.67&   -1.51&   -1.78&   -2.32&   -2.22&   -2.08&   -1.96&   -2.03\\\cline{2-13}
\(\chi=0.9\)&Med.      &-3.00& -2.83&   -2.69&   -0.61&   -1.49&   -1.61&   -1.97&   -1.88&   -1.87&   -1.73&  -1.90\\\cline{2-13}
&U. Qt.                        &-2.93& -2.79&   -2.65&   -0.57&   -1.48&   -1.46&   -1.73&   -1.70&   -1.63&   -1.45&   -1.72\\\hline

&Mean                       &-2.99&  -2.82&   -2.70&   -0.55&  -1.49&   -1.60&   -2.05&   -1.97&   -1.86&   -1.72&   -1.88\\\cline{2-13}
$q=0.3$&St. Dev.    &0.089&  0.095&  0.059&  0.062& 0.119&  0.317&  0.420&  0.440&  0.359&  0.392&  0.274\\\cline{2-13}
&L. Qt.                        &-3.06& -2.90&   -2.72&   -0.57&   -1.53&   -1.79&   -2.30&   -2.18&   -2.08&   -1.96&   -2.03\\\cline{2-13}
\(\chi=0.3\)&Med.      &-3.00& -2.83&   -2.70&   -0.54&   -1.48&   -1.56&   -1.95&   -1.87&   -1.86&   -1.72&  -1.92\\\cline{2-13}
&U. Qt.                        &-2.92& -2.76&   -2.68&   -0.52&   -1.47&   -1.44&   -1.73&   -1.68&   -1.61&   -1.45&   -1.72\\\hline

&Mean                       &-2.88&  -2.69&   -2.68&   -0.50&  -1.49&   -1.59&   -2.06&   -1.97&   -1.87&   -1.73&   -1.87\\\cline{2-13}
$q=0.3$&St. Dev.    &0.049&  0.053&  0.103&  0.099& 0.298&  0.354&  0.407&  0.456&  0.358&  0.393&  0.275\\\cline{2-13}
&L. Qt.                        &-2.93& -2.72&   -2.72&   -0.55&   -1.61&   -1.78&   -2.29&   -2.22&   -2.08&   -1.95&   -2.05\\\cline{2-13}
\(\chi=0.1\)&Med.      &-2.87& -2.70&   -2.70&   -0.51&   -1.53&   -1.63&   -1.97&   -1.89&   -1.87&   -1.72&  -1.90\\\cline{2-13}
&U. Qt.                        &-2.84& -2.67&   -2.67&   -0.49&   -1.43&   -1.46&   -1.71&   -1.66&   -1.63&   -1.45&   -1.72\\\hline
\end{tabular}
\caption{ As Table~\ref{tabFMErrIMBH}, but for an inspiralling IMBH with spin parameter \(q=0.3\). Results are quoted at a fixed SNR of 750.}
\label{tabFMErrIMBH03}
\end{table}

\begin{table}[thb]
\begin{tabular}{|c|c|c|c|c|c|c|c|c|c|c|c|c|}
\hline\multicolumn{2}{|c|}{}&\multicolumn{11}{c|}{Distribution of \(\log_{10}(\Delta X)\) in error, \(\Delta X\), for parameter \(X=\)}\\\cline{3-13}
\multicolumn{2}{|c|}{Model}&$\ln(m)$&$\ln(M)$&$q$&$\chi$&$p_0$&$\phi_0$&$\theta_S$&$\phi_S$&$\theta_K$&$\phi_K$&$\ln(D)$\\\hline
&Mean                       &-2.93&  -2.80&   -2.61&   -0.47&  -1.52&   -1.61&   -2.05&   -1.96&   -1.85&   -1.71&   -1.87\\\cline{2-13}
$q=0.1$&St. Dev.    &0.118&  0.133&  0.073&  0.073& 0.177&  0.332&  0.422&  0.418&  0.353&  0.416&  0.273\\\cline{2-13}
&L. Qt.                        &-2.99& -2.85&   -2.63&   -0.49&   -1.55&   -1.79&   -2.31&   -2.21&   -2.09&   -1.95&   -2.02\\\cline{2-13}
\(\chi=0.9\)&Med.      &-2.93& -2.79&   -2.59&   -0.45&   -1.50&   -1.63&   -1.94&   -1.87&   -1.85&   -1.71&  -1.90\\\cline{2-13}
&U. Qt.                        &-2.92& -2.73&   -2.57&   -0.41&   -1.47&   -1.49&   -1.72&   -1.67&   -1.61&   -1.45&   -1.69\\\hline

&Mean                       &-2.91&  -2.79&   -2.53&   -0.41&  -1.46&   -1.60&   -2.05&   -1.96&   -1.86&   -1.71&   -1.87\\\cline{2-13}
$q=0.1$&St. Dev.    &0.086&  0.089&  0.076&  0.071& 0.087&  0.321&  0.412&  0.404&  0.361&  0.395&  0.279\\\cline{2-13}
&L. Qt.                        &-2.95& -2.84&   -2.60&   -0.46&   -1.54&   -1.81&   -2.29&   -2.21&   -2.09&   -1.96&   -2.04\\\cline{2-13}
\(\chi=0.3\)&Med.      &-2.93& -2.81&   -2.55&   -0.42&   -1.47&   -1.56&   -1.96&   -1.89&   -1.85&   -1.70&  -1.92\\\cline{2-13}
&U. Qt.                        &-2.87& -2.73&   -2.46&   -0.35&   -1.38&   -1.42&   -1.72&   -1.68&   -1.62&   -1.46&   -1.71\\\hline

&Mean                       &-2.90&  -2.78&   -2.54&   -0.41&  -1.45&   -1.59&   -2.05&   -1.96&   -1.86&   -1.71&   -1.88\\\cline{2-13}
$q=0.3$&St. Dev.    &0.104&  0.109&  0.095&  0.095& 0.109&  0.321&  0.412&  0.412&  0.361&  0.365&  0.256\\\cline{2-13}
&L. Qt.                        &-2.96& -2.79&   -2.55&   -0.47&   -1.51&   -1.80&   -2.30&   -2.19&   -2.09&   -1.96&   -2.04\\\cline{2-13}
\(\chi=0.1\)&Med.      &-2.92& -2.76&   -2.56&   -0.42&   -1.48&   -1.62&   -1.96&   -1.88&   -1.86&   -1.71&  -1.91\\\cline{2-13}
&U. Qt.                        &-2.82& -2.68&   -2.47&   -0.32&   -1.37&   -1.44&   -1.72&   -1.69&   -1.62&   -1.42&   -1.70\\\hline
\end{tabular}
\caption{ As Table~\ref{tabFMErrIMBH}, but for an inspiralling IMBH with spin parameter \(q=0.1\). Results are quoted at a fixed SNR of 750. Note that as in Table~\ref{tabFMErrMMBHs}, when the central SMBH is slowly rotating,  the determination of the source parameters has no strong dependance on the the spin magnitude of the inspiralling IMBH.}
\label{tabFMErrIMBH01}
\end{table}
\clearpage

It is important to ask how much of an impact the additional terms we have included in the model have made to the parameter estimation results. The `best case scenario', i.e., IMRIs  with mass components \(10^4M_{\odot}+10^6M_{\odot}\) whose central SMBH is rapidly rotating, was not used for the original analysis reported in~\cite{smallbody}. We can evaluate the importance of the higher-order spin corrections by repeating the calculation for this new case using the previous model described in \cite{smallbody} to obtain parameter error measurements for IMBHs with spins \(\chi=0.9,\, 0.3,\, 0.1\). The results are shown in Table~\ref{tabFMErrIMBH2pn}.

\begin{table}[thb]
\begin{tabular}{|c|c|c|c|c|c|c|c|c|c|c|c|c|}
\hline\multicolumn{2}{|c|}{}&\multicolumn{11}{c|}{Distribution of \(\log_{10}(\Delta X)\) in error, \(\Delta X\), for parameter \(X=\)}\\\cline{3-13}
\multicolumn{2}{|c|}{Model}&$\ln(m)$&$\ln(M)$&$q$&$\chi$&$p_0$&$\phi_0$&$\theta_S$&$\phi_S$&$\theta_K$&$\phi_K$&$\ln(D)$\\\hline
&Mean                       &-2.83&  -2.65&   -3.71&   -0.87&  -1.20&   -1.53&   -2.01&   -1.95&   -1.81&   -1.67&   -1.84\\\cline{2-13}
$q=0.9$&St. Dev.    &0.072&  0.082&  0.071&  0.079& 0.086&  0.331&  0.488&  0.475&  0.391&  0.407&  0.283\\\cline{2-13}
&L. Qt.                        &-2.87& -2.72&   -3.78&   -0.93&   -1.26&   -1.74&   -2.26&   -2.21&   -2.05&   -1.90&   -2.01\\\cline{2-13}
\(\chi=0.9\)&Med.      &-2.82& -2.64&   -3.70&   -0.88&   -1.19&   -1.58&   -1.90&   -1.86&   -1.80&   -1.69&  -1.86\\\cline{2-13}
&U. Qt.                        &-2.75& -2.56&   -3.64&   -0.82&   -1.11&   -1.40&   -1.65&   -1.64&   -1.56&   -1.45&   -1.69\\\hline

&Mean                       &-2.81&  -2.62&   -3.69&   -0.82&  -1.15&   -1.52&   -2.00&   -1.93&   -1.81&   -1.66&   -1.83\\\cline{2-13}
$q=0.9$&St. Dev.    &0.087&  0.088&  0.082&  0.081& 0.088&  0.325&  0.443&  0.448&  0.375&  0.405&  0.290\\\cline{2-13}
&L. Qt.                        &-2.86& -2.69&   -3.74&   -0.87&   -1.21&   -1.71&   -2.26&   -2.18&   -2.03&   -1.89&   -2.00\\\cline{2-13}
\(\chi=0.3\)&Med.      &-2.80& -2.61&   -3.69&   -0.83&   -1.14&   -1.57&   -1.91&   -1.83&   -1.80&   -1.65&  -1.85\\\cline{2-13}
&U. Qt.                        &-2.73& -2.56&   -3.64&   -0.77&   -1.08&   -1.39&   -1.67&   -1.64&   -1.54&   -1.42&   -1.66\\\hline

&Mean                       &-2.79&  -2.60&   -3.68&   -0.78&  -1.13&   -1.50&   -1.99&   -1.92&   -1.81&   -1.65&   -1.83\\\cline{2-13}
$q=0.9$&St. Dev.    &0.086&  0.087&  0.079&  0.077& 0.091&  0.327&  0.420&  0.475&  0.373&  0.407&  0.294\\\cline{2-13}
&L. Qt.                        &-2.83& -2.66&   -3.72&   -0.84&   -1.19&   -1.70&   -2.23&   -2.17&   -2.02&   -1.90&   -1.99\\\cline{2-13}
\(\chi=0.1\)&Med.      &-2.78& -2.59&   -3.68&   -0.77&  -1.12&   -1.54&   -1.98&   -1.82&   -1.82&   -1.65&  -1.85\\\cline{2-13}
&U. Qt.                        &-2.72& -2.53&   -3.61&   -0.75&   -1.06&   -1.36&   -1.65&   -1.62&   -1.65&   -1.40&   -1.66\\\hline
\end{tabular}
\caption{ As Table~\ref{tabFMErrIMBH}, but including small body spin corrections up to 2PN order. Results are quoted at a fixed SNR of 1000.}
\label{tabFMErrIMBH2pn}
\end{table}

\begin{figure*}[ht]
\centerline{
\includegraphics[height=0.38\textwidth,angle=0,  clip]{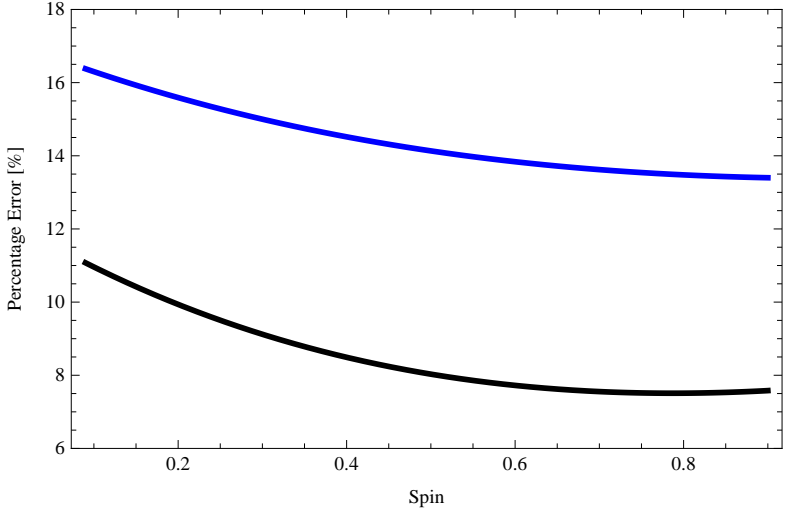}
}
\caption{The plot shows the accuracy with which GW observations may determine the spin parameter of   \(10^4M_{\odot}\) IMBHs that inspiral into  rapidly rotating \(10^6M_{\odot}\) SMBHs. We have used the median of a set of Monte Carlo simulations of Fisher Matrix errors to generate this plot.  We assume that these sources are observed at a fixed SNR\(=1000\). The upper line has been obtained using a model with small body spin corrections up to 2PN order~\cite{smallbody}. The bottom line includes all higher-order spin corrections, as described in this paper.}
\label{sperr}
\end{figure*}

A direct comparison between Tables~\ref{tabFMErrIMBH} and \ref{tabFMErrIMBH2pn} shows that the inclusion of higher-spin corrections in the waveform template increases our estimates of the accuracy with which GW observations might determine the parameters of these type of events.  In particular, we notice that our ability to determine the spin of the inspiralling IMBH is improved by nearly a factor of two, as shown in Figure~\ref{sperr}. 

There is an additional reason to include higher-order spin corrections in the waveform template for this kind of sources. As we discussed in \cite{smallbody}, a waveform template that includes small body spin corrections up to 2PN order has systematic errors whose magnitude is roughly a factor \(\sim 4\) larger than noise induced errors for intrinsic parameters. This initial analysis then suggested that in order to improve parameter error determination for IMRI sources, and enhance the reliability of these results, it was necessary to include higher-order spin corrections in the equations of motion to further reduce the ratio of systematic to noise induced errors. In the following Section we show that this approach is the right strategy to improve the accuracy of the waveform template.    

\section{Model--induced parameter errors}
\label{s6}

In the previous Section, we explored the capability of future low-frequency GW detectors, such as LISA, to determine the spin distribution of IMBHs through IMRI observations. We have shown that the inclusion of higher-order spin effects in the dynamics of compact binaries has an important effect, particularly for IMRIs with mass ratio \(\eta\gtrsim10^{-3}\).  In this Section, we explore the importance of including higher order PN corrections for parameter estimation.  The strategy to perform this analysis consists of building templates that include all or only part of the corrections derived in Section~\ref{s2}. 

The IMRI kludge model in Eq.~(\ref{omegacc}) includes corrections up to 2.5PN order. In this Section, a search template that includes all these corrections will be taken as the `true' waveform, \(h_{\rm {\bf GR}}\), whereas templates that include only some of the \(c_i\) coefficients of Eq.~\eqref{omCCone}, will be taken as approximate templates, \(h_{\rm {\bf AP}}\). For instance,  a template that includes corrections at 1.5PN order  includes the coefficients \(c_0, c_2, c_3\) and \(c_{3.1}\) only.  We explore the importance of including the corrections by changing the terms entering expression~\eqref{omegacc} for the frequency, but leaving intact the radiative part of the waveform model described by Eqs.~\eqref{new_Ldot}--\eqref{6}. This approach is consistent with how the coefficients in Eq.~(\ref{omegacc}) were derived from comparison to PN models. An understanding of systematic errors is particularly relevant for IMRI sources, because, as shown in Section~\ref{snrsec}, the typical SNRs range from a few hundred up to a thousand. Hence, it is possible that `model' or systematic errors, which are SNR independent, could dominate the total parameter-estimation error. 

In order to assess whether systematic errors dominate, we will compute the ratio \({\cal R}\) of the systematic error to the noise-induced error. If this analysis indicates that  \({\cal R}\sim 1\), then the statistical parameter error estimates that we presented in Section~\ref{s5} are a reliable indicator of the precision to which we would measure source parameters in practice. If \({\cal R}\gg1\) we would need to further  improve the accuracy of the model, either by including higher-order PN corrections for non-spinning binaries or by including second-order conservative corrections in order to accurately measure parameters. To estimate the magnitude of the systematic errors, we will use the formalism introduced in \cite{vallisneri}. At first order in the error \(\Delta\theta^i ({\bf n})\), we can identify two contributions. The first is due to noise in the detector, and the second can be related to the approximate nature of the waveform template, which introduces another source of error given by

\begin{eqnarray}
\Delta_\mathrm{th}\theta^i &=&  \Big(\Gamma^{-1}(\tbf)\Big)^{ij} \Big( \partial_j \hAP(\tbf) \Big| \hGR(\ttr) - \hAP(\ttr)\Big),
\label{38.2}
\end{eqnarray}

\noindent where \(\{h_{\rm GR}(\theta^i)\}\) and \(\{h_{\rm AP}(\theta^i)\}\) represent the two manifolds in the vector spaces of data streams corresponding to the `true' and approximate waveforms, respectively. Given the fact that: (i) we do not currently have `true' waveform templates at hand and we can only experimentally determine the  \(\hAP(\tbf)\) that is the best fit waveform for the data stream, \(\mathbf{s}= {\bf h}_\mathrm{GR}(\hat \theta^i) + {\bf n}\); and (ii) we are unsure about the error \(\Delta \theta \equiv \tbf - \ttr\), since we do not know $\hat{\theta}$, we make progress by replacing, at leading order, \(\hGR(\ttr) - \hAP(\ttr)\) by \(\hGR(\tbf) - \hAP(\tbf)\) to obtain

\begin{equation}
\label{39}
\Delta_\mathrm{th}\theta^i =  \Big(\Gamma^{-1}(\tbf)\Big)^{ij} \Big( \partial_j \hAP(\tbf) \Big| \hGR(\tbf) - \hAP(\tbf)\Big).
\end{equation}

\noindent The Fisher matrix is to be evaluated using the approximate waveforms \(\Gamma_{ij}(\tbf) \equiv ( \partial_i \hAP(\tbf)|\partial_j \hAP(\tbf))\).  As discussed in \cite{vallisneri}, this scheme works best when the waveform is rewritten using an amplitude-phase decomposition, i.e., 

\newcommand{\hla}{\mathbf{h}}
\newcommand{\tla}{\theta(\lambda)}
\newcommand{\ttl}{\theta_\mathrm{tr}(\lambda)}

\begin{equation}
\label{40}
\tilde h^{\alpha}(f) =  A^{\alpha}(f) e^{i \Psi^{\alpha}(f)} \,,
\end{equation}

\noindent where the amplitude \(A\) and phase \(\Psi\) are given by Eqs. (56)-(58) in \cite{smallbody}. 

\noindent In this form, the first order approximation of Eq.~\eqref{39} becomes
\begin{equation}
\label{42}
\Delta_\mathrm{th}\theta^i \approx
\big(\Gamma^{-1}(\tbf) \big)^{ij}  \Big( \underbrace{\big[ \Delta {\bf A} + i {\bf A} \Delta \boldsymbol{\Psi} \big] e^{i \boldsymbol{\Psi}}}_{\text{at}\,\theta} \Big| \,
  \partial_j \hAP(\tbf) \Big).
\end{equation}

In a previous study~\cite{smallbody}, we found  that IMRI waveforms that included conservative corrections up to 2PN order would have a ratio \(\cal{R}\)  of model to noise-induced errors of  \( {\cal R}  \gtrsim 4\). This result suggested that it was necessary to improve the IMRI model to find out whether the inclusion of higher-order spin effects and their associated conservative corrections could further reduce this ratio. Here we show that by using conservative corrections up to 2.5PN order we are able to reduce the ratio \(\cal{R}\)  down to  \({\cal R}\sim 1\) for the intrinsic parameters (mass, spin etc.). This would mean that the parameter estimation errors presented in the previous Section should be reliable. More importantly, this suggests that we will not need to go beyond 2.5PN to recover the source parameters accurately, even for strong IMRI sources. 

\begin{table}[thb]
\begin{tabular}{|c|c|c|c|c|c|c|c|c|c|c|c|c|}
\hline\multicolumn{2}{|c|}{}&\multicolumn{11}{c|}{\small{\(\log_{10}\) of the ratio \(\cal{R}\) of model to noise--induced error for parameter \(X=\)}}\\\cline{3-13}
\multicolumn{2}{|c|}{Model}&$\ln(m)$&$\ln(M)$&$q$&$\chi$&$p_0$&$\phi_0$&$\theta_S$&$\phi_S$&$\theta_K$&$\phi_K$&$\ln(D)$\\\hline
&Mean                                &0.38    &0.39   &0.76     &0.49   &0.39     &1.04    &0.93    &0.90   &1.04    &1.11    &1.10\\\cline{2-13}
2PN vs 1.5PN&St. Dev.  &0.719  &0.604  &0.490 &0.585  &0.594  &0.720  &0.715 &0.637 &0.799  &0.735  &0.816\\\cline{2-13}
&L. Qt.                                &-0.18    &-0.16  &0.56    &-0.11  &-0.18    &0.40    &0.38   &0.30    &0.39    &-0.02   &0.35\\\cline{2-13}
\(\chi=0.9\)&Med.             &0.47     &0.45   &0.82    &0.64     &0.51     &0.96    &1.00   &0.98    &1.08    &1.17    &1.11\\\cline{2-13}
&U. Qt.                               &0.90     &0.87   &1.04    &1.04     &0.84     &1.77    &1.54   &1.45    &1.78    &1.93    &1.94\\\hline

&Mean                                  &0.017    &0.016   &0.240     &0.038    &0.016     &0.334    &0.414     &0.351     &0.494    &0.450    &0.462\\\cline{2-13}
2.5PN vs 2.0PN&St. Dev. &0.609    &0.603   &0.563     &0.607     &0.603   &0.807   &0.755       &0.704      &0.835     &0.876  &0.848\\\cline{2-13}
&L. Qt.                                  &-0.285      &-0.257  &-0.050    &-0.482   &-0.257  &-0.340  &-0.232   &-0.380  &-0.271   &-0.411   &-0.396\\\cline{2-13}
\(\chi=0.9\)&Med.               &-0.011    &-0.012  &0.297     &0.052     &0.003   &0.259   &0.368       &0.407     &0.420     &0.463    &0.391\\\cline{2-13}
&U. Qt.                                  &0.403    &0.429   &0.377     &0.595     &0.420   &0.981    &1.210         &1.064      &1.445    &1.321   &1.404\\\hline

&Mean                                &0.50    &0.50  &0.77  &0.63  &0.50   &1.06  &0.97  &0.92  &1.11  &1.14  &1.10\\\cline{2-13}
2PN vs 1.5PN&St. Dev.  &0.708 &0.622 &0.405 &0.612 &0.621  &0.731 &0.632 &0.680 &0.701 &0.703 &0.795\\\cline{2-13}
&L. Qt.                                &0.02    &0.01  &0.58  &0.14  &0.03   &0.39  &0.38  &0.28  &0.34  &0.36  &0.30\\\cline{2-13}
\(\chi=0.3\)&Med.             &0.63    &0.63  &0.78  &0.77  &0.62   &1.00  &1.00  &0.97  &1.12  &1.16  &1.18\\\cline{2-13}
&U. Qt.                               &0.98    &0.98  &1.02  &1.11  &0.97   &1.73  &1.68  &1.60  &2.00  &2.01  &1.99\\\hline

&Mean                                  &0.081    &0.076   &0.398     &0.086    &0.064     &0.410    &0.497     &0.447     &0.575    &0.556    &0.588\\\cline{2-13}
2.5PN vs 2.0PN&St. Dev. &0.563    &0.493   &0.472     &0.630     &0.495   &0.807     &0.743       &0.687    &0.811     &0.854  &0.832\\\cline{2-13}
&L. Qt.                                  &-0.233    &-0.237  &0.344    &-0.407   &-0.211  &-0.263    &-0.206   &-0.194    &-0.206   &-0.189   &-0.149\\\cline{2-13}
\(\chi=0.3\)&Med.               &-0.041    &-0.045  &0.477     &-0.023   &-0.025   &0.366   &0.415       &0.435    &0.504     &0.557    &0.503\\\cline{2-13}
&U. Qt.                                  &0.441    &0.437   &0.539     &0.532     &0.416   &1.210    &1.341         &1.120      &1.412       &1.376   &1.423\\\hline
\end{tabular}
\caption{ Summary of Monte Carlo simulation results for the ratio of model errors to noise-induced errors, for IMRIs including an IMBH of mass  $\mu=5\times10^3M_{\odot}$ and a central SMBH with spin parameter \(q=0.9\).  We show the mean, standard deviation, median and quartiles of the distribution of the logarithm to base ten of the ratio for each parameter. Results are given for various comparisons, as indicated and described in the text. A comparison ``A vs B'' uses model A as the true waveform and model B as the search template. Note that the noise--induced errors are quoted at a fixed SNR\(=1000\).}
\label{tabFMRatMMBH}
\end{table}
\clearpage

Once the gravitational self-force results for inspirals in Kerr are available, we will be able to extend this model by using fully relativistic calculations. Nonetheless, we will still have to use perturbative and PN corrections for spin-spin and spin-orbit effects, since the scope of the current self--force program is focused on inspiraling objects that are non-spinning \cite{sago,sagoec,war,warleor}. The results we present below shed light on the level of accuracy that a waveform template should have to enable accurate parameter reconstruction for loud GW sources with two spinning components.

In Table~\ref{tabFMRatMMBH} and~\ref{tabFMRatMMBHI}, we present the ratio \(\cal{R}\) of model errors to noise--induced errors for the same systems considered in Section~\ref{s5} and for  two different comparisons, 2.5PN vs 2PN, and 2PN vs 1.5PN. The results show that the ratio \(\cal{R}\)  becomes smaller as the approximate waveform, \(\hAP\), becomes closer to the ``true'' waveform, \(\hGR\), and more importantly, that for the  2.5PN \(\rightarrow\) 2PN comparison \({\cal R}\sim 1\) for the intrinsic parameters. The error ratio is somewhat bigger for the extrinsic parameters, \({\cal R}\sim 4\). The fact that the error ratio has improved for the intrinsic parameters but not the extrinsic parameters is perhaps not surprising. Information about the intrinsic parameters comes primarily from the waveform phasing while that on the extrinsic parameters comes from amplitude modulations. The corrections we have added into the model in this paper are corrections to the phase evolution and therefore we would expect them to primarily impact errors in the intrinsic parameters, which is what we see. This suggests that we need to include conservative corrections up to 2.5PN order in IMRI search templates in order to recover accurate estimates of the intrinsic parameters, but that with templates of this order, model errors should no longer be a limiting factor.

As discussed in Section~\ref{s5}, IMRIs with mass components \(10^{4} M_{\odot}+ 10^{6}M_{\odot}\) may provide the best chance to determine the spin distribution of IMBHs that fall into rapidly rotating SMBHs, even if the IMBH is slowly rotating. Table~\ref{tabFMRatMMBHI} shows that for these sources, the ratio  \(\cal{R}\)  becomes significantly smaller for the 2.5PN \(\rightarrow\) 2PN comparison and that \({\cal R} \sim 1\) for the intrinsic parameters as in the previous case, while again ${\cal R}\sim 4$ for the extrinsic parameters\footnote{Strictly speaking, \({\cal R} \sim 1\)  for all the intrinsic parameters, except for the central SMBH spin parameter, for which \({\cal R}\lesssim 2\).}. We expect to do better for systems of higher mass ratio, because the corrections that we are omitting in this comparison are proportional to the mass-ratio. Cutler \& Vallisneri~\cite{vallisneri} also presented results for systems with component masses $10^4M_\odot+10^6M_{\odot}$, although their results were for binaries in which neither component had spin and they compared a consistent $3$PN approximate waveform to a ``true'' $3.5$PN waveform. They found that the error ratios were larger for intrinsic parameters  ${\cal R} \sim 10$ than for the extrinsic parameters ${\cal R} \sim$ a few. The results we have here are in fact fairly consistent with their findings, in the sense that both the estimated error ratio for the extrinsic parameters and the magnitude of the systematic errors for the intrinsic parameters are very consistent between the two studies. The apparent discrepancy arises almost entirely from a difference in the magnitude of the statistical errors for the intrinsic parameters, which are one to two orders of magnitude smaller in their work. The inclusion of spin in the waveform model introduces correlations that tend to decrease the precision with which the system parameters can be measured and this is the explanation for the difference between the current work and that previous study.

\begin{table}[thb]
\begin{tabular}{|c|c|c|c|c|c|c|c|c|c|c|c|c|}
\hline\multicolumn{2}{|c|}{}&\multicolumn{11}{c|}{\small{\(\log_{10}\) of the ratio \(\cal{R}\) of model to noise--induced error for parameter \(X=\)}}\\\cline{3-13}
\multicolumn{2}{|c|}{Model}&$\ln(m)$&$\ln(M)$&$q$&$\chi$&$p_0$&$\phi_0$&$\theta_S$&$\phi_S$&$\theta_K$&$\phi_K$&$\ln(D)$\\\hline
&Mean                                &0.44    &0.49   &0.90     &0.51   &0.45     &1.12    &0.98    &0.94   &1.08    &1.15    &1.08\\\cline{2-13}
2PN vs 1.5PN&St. Dev.  &0.722  &0.618  &0.519 &0.542  &0.613  &0.745  &0.723 &0.678 &0.781  &0.637  &0.821\\\cline{2-13}
&L. Qt.                                &-0.05    &-0.08  &0.72    &-0.09  &-0.07    &1.50    &0.27   &0.33    &0.41    &0.34   &0.38\\\cline{2-13}
\(\chi=0.9\)&Med.             &0.54     &0.51   &0.98    &0.66     &0.53     &0.91    &1.00   &1.01    &1.11    &1.04    &1.09\\\cline{2-13}
&U. Qt.                               &0.96     &0.93   &1.06    &1.11     &0.96     &1.62    &1.65   &1.51    &1.95    &1.95    &1.96\\\hline

&Mean                                  &0.00    &0.014   &0.345     &0.009       &0.006      &0.380    &0.392     &0.349     &0.504    &0.469    &0.496\\\cline{2-13}
2.5PN vs 2.0PN&St. Dev. &0.569    &0.550   &0.475     &0.708     &0.541      &0.722   &0.703      &0.802      &0.823     &0.735  &0.705\\\cline{2-13}
&L. Qt.                                  &-0.263    &-0.230  &0.253    &-0.439   &-0.234     &-0.288  &-0.277    &-0.321    &-0.261   &-0.328   &-0.325\\\cline{2-13}
\(\chi=0.9\)&Med.               &-0.119    &-0.085  &0.375     &-0.072   &-0.098    &0.302   &0.350       &0.340     &0.401     &0.421    &0.403\\\cline{2-13}
&U. Qt.                                  &0.314    &0.349   &0.433     &0.505     &0.351     &0.952    &1.150       &0.949     &1.436    &1.245   &1.370\\\hline

&Mean                                &0.57    &0.58  &0.93  &0.63  &0.59       &1.07  &1.01  &0.99      &1.17  &1.18  &1.19\\\cline{2-13}
2PN vs 1.5PN&St. Dev.  &0.715 &0.618 &0.466 &0.612 &0.718  &0.743 &0.583 &0.508 &0.683 &0.694 &0.735\\\cline{2-13}
&L. Qt.                                &0.06    &0.08  &0.75  &0.14  &0.08        &0.38  &0.37  &0.37      &0.45  &0.47  &0.44\\\cline{2-13}
\(\chi=0.3\)&Med.             &0.71    &0.69  &1.01  &0.77  &0.68         &0.99  &1.06  &1.05      &1.14  &1.17  &1.22\\\cline{2-13}
&U. Qt.                               &1.07    &1.09  &1.14  &1.11  &1.10         &1.74  &1.74  &1.62      &2.03  &2.09  &2.01\\\hline

&Mean                                  &0.075    &0.069   &0.357     &0.058    &0.055     &0.298    &0.425     &0.364     &0.527    &0.457    &0.517\\\cline{2-13}
2.5PN vs 2.0PN&St. Dev. &0.556    &0.554   &0.470     &0.631     &0.507    &0.781     &0.737       &0.539    &0.763     &0.703  &0.704\\\cline{2-13}
&L. Qt.                                  &-0.286    &-0.276  &0.215    &-0.467   &-0.289   &-0.418    &-0.318   &-0.369    &-0.315   &-0.299   &-0.305\\\cline{2-13}
\(\chi=0.3\)&Med.               &0.050    &0.070   &0.458     &0.099    &0.061     &0.235    &0.355       &0.362    &0.516     &0.513    &0.502\\\cline{2-13}
&U. Qt.                                  &0.513    &0.492   &0.579    &0.550     &0.444    &0.918    &1.339         &1.095    &1.324     &1.337   &1.411\\\hline
\end{tabular}
\caption{ As Table~\ref{tabFMRatMMBH}, but for IMRI  sources with mass components \(10^{4} M_{\odot}+ 10^{6}M_{\odot}\). Noise--induced errors are quoted at a fixed SNR\(=1000\).}
\label{tabFMRatMMBHI}
\end{table}

In summary, we have demonstrated that  a  waveform template that includes all currently available higher-order spin effects and conservative PN corrections for spin-spin and spin-orbit effects might be sufficient to enable us to accurately reconstruct the parameters of loud sources, since the relative importance of the \(2.5{\rm PN}\rightarrow2{\rm PN}\) change is negligible, even for the loudest IMRI sources considered here. 

\clearpage
\section{Conclusions}
\label{s7}

This paper is the second in a series that sheds light on the importance of including small body spin correction in the modeling of extreme and intermediate mass ratio inspirals. In the first paper~\cite{smallbody}, we showed that these effects are important for IMRI sources with mass ratios \(\eta\gtrsim10^{-3}\), and that future low-frequency GW detectors might provide an accurate measurement of the spin of IMBHs that inspiral into SMBHs, as long as both compact objects are rapidly rotating. In this paper we have improved our initial analysis in several important ways. First, we improved the waveform model by including the best information currently available  from perturbative \cite{tanaka} and PN results \cite{buoII,blanchet,buoerr1,buoerr2} to amend the equations of motion that describe the inspiral evolution of spinning compact objects into Kerr black holes \cite{maeda}.  We showed how to implement first order conservative corrections for spin-spin and spin-orbit effects  at 2.5PN order to compute the evolution of the inspiralling object's orbital frequency. Furthermore, we generated the trajectory of the inspiralling CO  using the most accurate prescriptions available for the fluxes of energy and angular momentum which include higher-order fits to Teukolsky-based evolutions \cite{improved} and perturbative results at 2.5PN order \cite{tanaka} for spin-spin and spin-orbit couplings.

We then used this improved model to refine our understanding of the capability of future low-frequency detectors to measure the spin distribution of IMBHs that inspiral into rapidly rotating $(q=0.9)$ SMBHs. Our first important finding was that for IMRI sources with mass components \(5\times10^{3}M_{\odot} + 10^{6}M_{\odot}\), GW observations at a fixed SNR=\(1000\) could determine the spin  of the IMBH  to a fractional accuracy of  \(\sim10\%\) for IMBHs with moderate to rapid spins. This is an important result in light of the current understanding on the formation channels of IMBHs. It is currently accepted that moderate spins would favour formation through mergers with comparable mass BHs, while rapid spins would favour growth through accretion \cite{seoane}. Hence, these results indicate that future low-frequency GW observations will provide valuable information on IMBH growth. 

Furthermore, we have shown that for IMRI sources with mass ratios \(\eta\sim 10^{-2}\), higher-order spin effects must be included in search templates for these events. This is because a search template that includes these corrections should allow us, through GW observations, to measure the spin of IMBHs that inspiral into rapidly rotating SMBHs to fractional accuracies  better than  \(\sim10\%\)   at a fixed SNR\(=1000\). Our results also suggest that when the central SMBH is moderately rotating, GW observations would enable us to constrain the spin of inspiralling IMBHs to \(\sim 30\%\), at SNR\(=750\).  Thus, events with \(\eta\sim 10^{-2}\) and rapidly rotating central SMBHs can be considered as ``golden'' IMRI sources, since they will provide a wealth of information on the existence, astrophysical properties, and physical processes that lead to the formation of IMBHs.

The second part of our analysis was an important check on our results.  We showed in Section~\ref{snrsec} that IMRI sources are intrinsically loud, which means that systematic errors, which are SNR independent, could dominate the total parameter error budget and limit our ability to obtain accurate parameter error estimates. If systematic errors were larger than noise-induced errors, our parameter errors estimates would not be trustworthy. In Section~\ref{s6}, we estimated the magnitude of systematic errors that would arise when part of the higher order corrections in the search templates were omitted. These results showed that the relative error when using a \(2{\rm PN}\) template to detect a \(2.5{\rm PN}\) signal is small at fixed SNR, for all the systems considered in our analysis.  Indeed, for the 2.5PN vs 2PN comparison, the ratio \({\cal{R}} \) of the model errors to the noise--induced errors is of order \({\cal{R}} \lesssim 1\) for the intrinsic parameters, even for the loudest sources detected with SNR of $1000$. This analysis indicated that by including higher-order spin effects, and their associated conservative corrections in search templates up to 2.5PN order would help us reduce the ratio of model to noise-induced errors for intrinsic parameters by a factor of \(\sim 4\), as compared to search templates that include these corrections up to 2PN order \cite{smallbody}. Hence, we conclude that to accurately reconstruct the parameters of strong IMRI sources, we might not need to go beyond 2.5PN order. This finding complements the study carried out in \cite{vallisneri}, where the authors showed that model errors could be a limiting factor in recovering the parameters of loud sources in which the components were non-spinning. In the current paper we have focussed on IMRI sources in which both components have spin and our waveform model is based on accurate equations of motion for circular-equatorial spinning BH binary systems, allowing it to be more readily applied in the strong-field. 

There is a second type of IMRI that could enhance our understanding of the astrophysical properties of IMBHs --- LIGO IMRIs, in which  a stellar-mass object ($\sim 1M_\odot$) falls into an IMBH ($\sim 1000M_\odot$). These could be detected by future ground-based detectors, such as Advanced LIGO \cite{network} or the Einstein Telescope~\cite{Freise:2009}. Previous studies have explored the capability of the Einstein Telescope to determine the astrophysical properties of these sources when the inspiralling black hole is not spinning~\cite{firstpaper,thirdpaper}. In light of  the findings presented in this paper, we see the need to improve this model by including small body spin effects, and recent results from fully accurate numerical relativity simulations of systems with mass ratios \(\eta\sim 10^{-2}\)~\cite{carlos}.  The development of these search templates will improve the scientific payoffs that will be obtained through GW observations in the Advanced GW detector era.

\section*{Acknowledgments}
This work was supported by NSF grants PHY-0847611 and PHY-0854812. JG's work is supported by the Royal Society.  DB and EH would also like to thank the Research Corporation for Science Advancement and the Cottrell Scholars program for support. The Monte Carlo simulations described in this paper were performed using the  Syracuse University Gravitation and Relativity  (SUGAR),  which is supported by NSF grants PHY-1040231, PHY-0600953 and PHY-1104371.

\bibliography{references}

\end{document}